\documentclass{article}
\usepackage[english]{babel}
\usepackage{xcolor}
\usepackage{amsmath}
\usepackage{amsfonts}
\usepackage{amsthm}
\usepackage{amssymb}
\usepackage{calrsfs}
\usepackage{graphicx}
\usepackage{bussproofs}
\usepackage{proof}
\usepackage{latexsym}
\usepackage[ND,SEQ]{prftree}
\usepackage{multicol, caption}
\usepackage{setspace}
\usepackage[nottoc,numbib]{tocbibind}
\usepackage{pdflscape}
\usepackage{stmaryrd}
\usepackage{wasysym}
\usepackage[natbibapa]{apacite}
\usepackage[titletoc,title]{appendix}
\newtheoremstyle{break}
  {\topsep}{\topsep}%
  {\itshape}{}%
  {\bfseries}{}%
  {\newline}{}%
\theoremstyle{break}
\newtheorem{theorem}{Theorem}[section]

\newtheorem{definition}{Definition}
\newtheorem{lemma}{Lemma}

\begin{document}

\title{Meaning and identity of proofs in a bilateralist setting: \\{\large A two-sorted typed lambda-calculus for proofs and refutations}}


\author{Sara Ayhan\thanks{I would like to thank Heinrich Wansing, David Ripley and Lloyd Humberstone for kindly dedicating their time to read earlier drafts of this paper, meet and discuss the ideas with me extensively. Thanks also go to two anonymous referees for their very thorough and helpful reports and to the UCL PPLV group and the Buenos Aires Logic Group for inviting me to give a talk about this paper and for the following discussions and comments.}
\\
{\small Ruhr University Bochum, Institute of Philosophy I}\\
{\small sara.ayhan@rub.de}}
\date{}

\maketitle
\begin{abstract}
\noindent In this paper I will develop a $\lambda$-term calculus, $\lambda^{2Int}$, for a bi-intuitionistic logic and discuss its implications for the notions of sense and denotation of derivations in a bilateralist setting.
Thus, I will use the Curry-Howard correspondence, which has been well-established between the simply typed $\lambda$-calculus and natural deduction systems for intuitionistic logic, and apply it to a bilateralist proof system displaying two derivability relations, one for proving and one for refuting.
The basis will be the natural deduction system of Wansing's bi-intuitionistic logic \texttt{2Int}, which I will turn into a term-annotated form.
Therefore, we need a type theory that extends to a two-sorted typed $\lambda$-calculus.
I will present such a term-annotated proof system for \texttt{2Int} and prove a Dualization Theorem relating proofs and refutations in this system.
On the basis of these formal results I will argue that this gives us interesting insights into questions about sense and denotation as well as synonymy and identity of proofs from a bilateralist point of view.
\end{abstract}
\section{Introduction}
In this paper I want to discuss the question how proofs and refutations can be related concerning their identity and meaning from a bilateralist perspective in proof-theoretic semantics.
Taking the perspective of proof-theoretic semantics means that I will assume that the meaning of the logical connectives is given by the rules of inference governing them in our underlying proof system.
The perspective of \emph{bilateralist} proof-theoretic semantics means that I will not only consider rules giving proof conditions for the connectives but also rules giving refutation conditions as determining the overall meaning of a connective.
Therefore, the bi-intuitionistc logic \texttt{2Int} is especially well-suited since it implements bilateralism deeply in the proof system in that it comprises two derivability relations, one displaying provability and the other displaying refutability \citep{Wansing2017}.

\noindent Bilateralism is situated in the area of proof-theoretic semantics (PTS).\footnote{For more detailed introductions and overviews on bilateralism and proof-theoretic semantics, see \citep{Ayhan2023I} and \citep{sep-proof-theoretic-semantics, Francez2015} respectively.}
In PTS, which can be seen as belonging to the broader approach of inferentialism, the meaning of logical connectives is determined by the rules of inference that govern their use in proofs. 
Thus, PTS opposes the traditional conception of model-theoretic semantics which takes the meaning of logical connectives to be given in terms of truth tables, first-order models, etc. 
Bilateralism, in a nutshell, is the view that dual concepts like truth and falsity, assertion and denial, or, in our context, proof and refutation should each be considered equally important and taken as primitive concepts, i.e., not,  like it is traditionally done, to see the latter as reducible to the former concepts. 
Although this position has been voiced in preceding papers (e.g., \cite{Price1983, Smiley1996, Humberstone2000}), the term ``bilateralism'' was coined by Rumfitt in his seminal paper \citeyearpar{Rumfitt2000} in which he argues for a ‘bilateral’ proof system of natural deduction with introduction and elimination rules determining not only the assertion conditions for formulas containing the connective in question but also the denial conditions.
Since then, different forms of implementing bilateralist ideas with respect to proof theory have evolved (see, e.g., contributions in \cite{Ayhan2023SI}), one of these being what will be considered in this paper, namely taking the concepts of proofs and refutations (instead of the speech acts of assertion and denial) to be on a par.

To approach the question of this paper, I will develop a $\lambda$-term calculus, $\lambda^{2Int}$, with which I will annotate Wansing's natural deduction system for \texttt{2Int} \citeyearpar{Wansing2016a, Wansing2017}.
Thus, I will use the Curry-Howard correspondence, which has been well-established between the simply typed $\lambda$-calculus and natural deduction systems for intuitionistic logic, and apply it to a bilateralist proof system.
Therefore, we need a type theory that extends to a two-sorted typed $\lambda$-calculus similar to what \cite{Wansing2016b} presents for the bi-connexive logic \texttt{2C}.
He uses a type theory \`{a} la Church, though, while I will introduce a Curry-style type theory.\footnote{This is not to say that this is the only difference, though. \texttt{2C} extends the language of \texttt{2Int} in that strong negation is added to the vocabulary. Apart from that, the systems differ with respect to their (co-)implications since \texttt{2C} comprises the connexive implication that can be found in the logic \texttt{C} \citep{Wansing2005} and accordingly a dualized version of that as co-implication. Thus, the resulting type systems also crucially differ in these respects.}
I will present such a term-annotated proof system for \texttt{2Int} and prove some properties and results for it (Section 2.1), most importantly for this paper a Dualization Theorem relating proofs and refutations in this system (Section 2.2).
On the basis of these formal results I will discuss its use and implications for the notions of sense and denotation of derivations in a bilateralist setting (Section 3). 
Specifically, I will argue that such a system makes explicit that proofs and refutations should be identified on the level of denotation but not on the level of sense, i.e., that the derivations refer to the same underlying object but are not synonymous.

\section{A two-sorted typed $\lambda$-calculus for \texttt{2Int}: $\lambda^{2Int}$} 
\subsection{The term-annotated natural deduction calculus \texttt{N2Int}$_\lambda$ and some results for the system}
Let \texttt{Prop} be a countably infinite set of atomic formulas. Elements from \texttt{Prop} will be denoted $\rho$, $\sigma$, $\tau$, $\rho_1$, $\rho_2$ ... etc. Formulas generated from \texttt{Prop} will be denoted $A, B, C, A_1, A_2$, ... etc.
We use $\Gamma$, $\Delta$,... for sets of formulas.
The concatenation $\Gamma$, $A$ stands for $\Gamma \cup \left\{A\right\}$.

\noindent The language $\mathcal{L}_{2Int}$ of \texttt{2Int}, as given by Wansing, is defined in Backus-Naur form as follows:
\begin{center}

$A ::= \rho \mid \bot \mid \top \mid (A \wedge A) \mid (A \vee A) \mid (A \rightarrow A) \mid (A \Yleft A)$.
\end{center}

\noindent In \citep{Wansing2017} a natural deduction system for \texttt{2Int} is given and a normal form theorem proven for it.
The system \texttt{N2Int} comprises besides the usual introduction and elimination rules for intuitionistic logic (henceforth: the \emph{proof rules}, indicated by using single lines) also rules that allow us to introduce and eliminate our connectives into and from dual proofs.\footnote{Especially in the later sections when I will discuss more philosophical issues, I will often use ``refutations'' instead of ``dual proofs''. The latter is the terminologically stricter expression, which is appropriate when we speak about the proof system, but it expresses essentially the same concept as the former.}
These so-called \emph{dual proof rules} (indicated by using double lines) are obtained by a dualization of the proof rules and having these two independent sets of rules is exactly what reflects the bilateralism of the proof system.\footnote{See \citep[p. 32-34]{Wansing2017}, for the description and the rules of the calculus. Apart from the term annotations, our presentation of \texttt{N2Int} given below differs in two minor aspects from the presentation in \citep{Wansing2017}: Firstly, I include explicit introduction rules for $\bot$ and $\top$, and secondly, I use dashed lines in four of the rules indicating that the conclusion can be obtained either by a proof or by a dual proof. These versions of the rules are derivable in Wansing's original \texttt{N2Int}, though.}
Also, a distinction is drawn in the premises between \emph{assumptions} (taken to be verified) and \emph{counterassumptions} (taken to be falsified).
This is indicated by an ordered pair $(\Gamma; \Delta)$ (with $\Gamma$ and $\Delta$ being finite, possibly empty sets) of assumptions ($\Gamma$) and counterassumptions ($\Delta$), together called the \emph{basis} of a derivation.
Single square brackets denote a possible discharge of assumptions, while double square brackets denote a possible discharge of counterassumptions.
If there is a derivation of $A$ from a (possibly empty) basis $(\Gamma; \Delta)$ whose last inference step is constituted by a proof rule, this will be indicated by $(\Gamma; \Delta) \Rightarrow^{+} A$. 
If there is a derivation of $A$ from a (possibly empty) basis $(\Gamma; \Delta)$ whose last inference step is constituted by a dual proof rule, this will be indicated by $(\Gamma; \Delta) \Rightarrow^{-} A$. 
We assume that if  $(\Gamma; \Delta) \Rightarrow^{*} A$, $\Gamma \subseteq \Gamma'$ and $\Delta \subseteq \Delta'$, then $(\Gamma'; \Delta') \Rightarrow^{*} A$.

\noindent Whenever the superscript $^{*}$ is used with a symbol, this is to indicate that  the superscript can be either $+$ or $-$ (called \emph{polarities}).
When $^{*}$ is used multiple times within a symbol, this is meant to always denote the same polarity. 
In contrast, when $^{\dagger}$ is used next to $^{*}$ in a symbol this means that it can - but does not have to - be of another polarity (yet again multiple $^{\dagger}$ denote the same polarity), i.e., for example $\texttt{case}~ r^{*}\{x^{*}.t^{\dagger} | y^{*}.s^{\dagger}\}^{\dagger}$ could either stand for $\texttt{case}~ r^{+}\{x^{+}.t^{+} | y^{+}.s^{+}\}^{+}$, $\texttt{case}~ r^{-}\{x^{-}.t^{-} | y^{-}.s^{-}\}^{-}$,  $\texttt{case}~ r^{+}\{x^{+}.t^{-} | y^{+}.s^{-}\}^{-}$, or $\texttt{case}~ r^{-}\{x^{-}.t^{+} | y^{-}.s^{+}\}^{+}$ but not for, e.g., $\texttt{case}~ r^{+}\{x^{+}.t^{+} | y^{+}.s^{-}\}^{-}$.
Furthermore, we use `$\equiv$' to denote \textit{syntactic identity} between terms or types.

\noindent With $\Yleft$, which we call ``co-implication'', we have a connective in this language which acts as a dual to implication.\footnote{Sometimes also called ``pseudo-difference'', e.g. in \cite{Rauszer}, or ``subtraction'', e.g. in \cite{Restall1997}, and used with different symbols.}
With that we are in the realm of so-called \emph{bi-intuitionistic logic}, which is a conservative extension of intuitionistic logic by co-implication.
Note that there is also a use of `bi-intuitionistic logic' in the literature to refer to a specific system, namely \texttt{BiInt}, also called `Heyting-Brouwer logic'.
Co-implication is there to be understood to internalize the preservation of non-truth from the conclusion to the premises in a valid inference. 
The system \texttt{2Int}, which is treated here, uses the same language as \texttt{BiInt}, but the meaning of co-implication differs in that it internalizes the preservation of falsity from the premises to the conclusion in a dually valid inference.\footnote{See \cite[][p. 30ff.]{Wansing2016a, Wansing2016c, Wansing2017} for a thorough comparison of co-implication in these two systems both w.r.t. their model-theoretic as well as their proof-theoretic interpretation.}
The systems also crucially differ in that \texttt{BiInt} does not work with two consequence relations.
One resulting difference that seems worth mentioning is that \texttt{2Int} enjoys both the \emph{disjunction property}, saying that if $A \vee B$ is provable, either $A$ is provable or $B$ is provable, and the \emph{dual conjunction property}, saying that if $A \wedge B$ is refutable, either $A$ is refutable or $B$ is refutable, while \texttt{BiInt} has neither \citep{PintoUustalu2018}.

\noindent From the viewpoint of bilateralism, i.e., considering falsificationism being on a par with verificationism, it is quite natural to extend our language by a connective for co-implication. The reason for this is that co-implication plays the same role in falsificationism as implication in verificationism: Both can be understood to express a concept of entailment in the object language.
If we expect $\Rightarrow^{+}$ to capture verification from the premises to the conclusion in a valid inference and $\Rightarrow^{-}$ to capture falsification from the premises to the conclusion in a dually valid inference, then, just like implication internalizes provability in that we have in our system $(A; \emptyset) \Rightarrow^{+} B$ iff $(\emptyset; \emptyset) \Rightarrow^{+} A \rightarrow B$, likewise co-implication internalizes dual provability in that we have $(\emptyset;A) \Rightarrow^{-} B$ iff $(\emptyset; \emptyset) \Rightarrow^{-} B \Yleft A$.

\vspace{-0.1cm}
\begin{definition}
 The set of \emph{type symbols} (or just \emph{types}) is the set of all formulas of $\mathcal{L}_{2Int}$. Let \texttt{Var$_{2Int}$} be a countably infinite set of two-sorted term variables. Elements from \texttt{Var$_{2Int}$} will be denoted $x^{*}$, $y^{*}$, $z^{*}$, $x_1^{*}$, $x_2^{*}$ ... etc. The two-sorted terms generated from \texttt{Var$_{2Int}$} will be denoted $t^{*}, r^{*}, s^{*}, t_1^{*}, t_2^{*}$, ... etc. The set \texttt{Term$_{2Int}$} can be defined in Backus-Naur form as follows:\\
$t ::= x^{*} \mid \texttt{top} ^{+} \mid \texttt{bot} ^{-} \mid abort(t^{*})^{\dagger} \mid \langle t^{*},t^{*}\rangle^{*} \mid fst(t^{*})^{*} \mid snd(t^{*})^{*} \mid inl(t^{*})^{*} \mid inr(t^{*})^{*} \mid \texttt{case}~ t^{*}\{x^{*}.t^{\dagger} | x^{*}.t^{\dagger}\}^{\dagger} \mid (\lambda x^{*}.t^{*})^{*} \mid App(t^{*},t^{*})^{*} \mid \{ t^{+},t^{-} \}^{*} \mid \pi_1(t^{*})^{\dagger} \mid \pi_2(t^{*})^{\dagger}$.\footnote{The question why we need polarities for the terms at all was raised by an anonymous referee. It will indeed be important both for philosophical reasons (see Section 3.2) as well as for reasons of having unique principal types in our system. For the latter, note that without the polarities the terms annotating the conclusion of a proof rule of one connective would be identical with the terms annotating the conclusion of the corresponding dual proof rule of its dual connective.}\\

\end{definition}

\begin{definition}
A \emph{(type assignment) statement} is of the form $t: A$ with term $t$ being the \emph{subject} and type $A$ the \emph{predicate} of the statement. It is read ``term $t$ is of type $A$'' or, in the `proof-reading', ``$t$ is a proof of formula $A$". In a basis $(\Gamma; \Delta)$ $\Gamma$ and $\Delta$ are now to be understood as sets of type assignment statements.
\end{definition}

\noindent We are thus using a type-system \textit{\`{a} la Curry}, in which the terms are not typed, in the sense that the types are part of the term's syntactic structure, but are \textit{assigned} types. 
Substitution for terms is expressed by $t[s/x]$, meaning that in term $t$ every free occurrence of $x$ is substituted by $s$.
The usual capture-avoiding requirements for variable substitution are to be observed.

\begin{definition}

We write that there is a \emph{derivation} $(\Gamma; \Delta) \Rightarrow_{N2Int_\lambda}^{*} t^{*}:A$ to express that $t^{*}: A$ is derivable in \texttt{N2Int}$_\lambda$ from $(\Gamma; \Delta)$ if:
\begin{itemize}
\item $t^{*}=x^+$ and $x^+:A \in \Gamma$, or
\item $t^*=x^-$ and $x^-:A \in \Delta$, or
\item $t^*:A$ can be produced as the conclusion from the premises $(\Gamma;\Delta)$ (with $\Gamma'\cup\Gamma''\cup\Gamma'''=\Gamma$ and $\Delta' \cup \Delta''\cup\Delta'''=\Delta$) according to the following rules:\footnote{The subscript of $\Rightarrow$ will be omitted henceforth unless there is a possibility for confusion.}
\end{itemize}

\end{definition}

\vspace{0.1cm}
\textbf{\texttt{N2Int}$_\lambda$}

\begin{center}

\quad \hspace{-0.6cm} 
\infer=[\scriptstyle\bot I^{d}]{\texttt{bot} ^{-}:\bot}{\infer*{}{(\Gamma;\Delta)}}
\quad \quad 
\deduce[\scriptstyle---------\bot E]{abort(t^{+})^{*}:A}{\infer{t^{+}:\bot}{\infer*{}{(\Gamma;\Delta)}}}
\quad  \quad \quad  
\infer[\scriptstyle\top I]{\texttt{top} ^{+}:\top}{\infer*{}{(\Gamma;\Delta)}}
\quad \quad 
\deduce[\scriptstyle---------\top E^{d}]{abort(t^{-})^{*}:A}{
	\infer={t^{-}:\top}{\infer*{}{(\Gamma;\Delta)}}}
\end{center}

\begin{center}
\vspace{0.2cm}
\quad \hspace{-0.8cm}  
\infer[\scriptstyle\wedge I]{\langle s^{+},t^{+}\rangle^{+}: A \wedge B }{\;\infer{s^{+}: A}{\infer*{}{(\Gamma' ; \Delta')}} \quad \quad \infer{t^{+}: B}{\infer*{}{(\Gamma'';\Delta'')}}}
\quad \quad \quad  
\infer[\scriptstyle\wedge E_{1}]{fst(t^{+})^{+}: A}{\infer{t^{+}:A \wedge B}{\infer*{}{(\Gamma;\Delta)}}}
\quad \quad \quad
\infer[\scriptstyle\wedge E_{2}]{snd(t^{+})^{+}:B}{\infer{t^{+}:A \wedge B}{\infer*{}{(\Gamma;\Delta)}}}
\end{center}

\begin{center}
\vspace{0.2cm}
\quad \hspace{-0.8cm}  
\infer=[\scriptstyle\wedge I^{d}_{1}]{inl(t^{-})^{-}: A \wedge B}{\infer={t^{-}:A}{\infer*{}{(\Gamma;\Delta)}}}
\quad \quad \quad
\infer=[\scriptstyle\wedge I^{d}_{2}]{inr(t^{-})^{-}:A \wedge B}{\infer={t^{-}:B}{\infer*{}{(\Gamma;\Delta)}}}
\end{center}

\begin{center}
\vspace{0.2cm}
\hspace{-0.4cm}  
\deduce[\scriptstyle-----------------------------------\wedge E^{d}]{\texttt{case}~ r^{-}\{x^{-}.s^{*} | y^{-}.t^{*}\}^{*}: C}{\infer={r^{-}: A \wedge B}{\infer*{}{(\Gamma';\Delta')}} \quad \deduce[\scriptstyle---]{s^{*}:C}{\infer*{}{(\Gamma'';\Delta'', \llbracket x^{-}:A\rrbracket)}} \quad \deduce[\scriptstyle---]{t^{*}:C}{\infer*{}{(\Gamma'''; \Delta''', \llbracket y^{-}:B\rrbracket)}}}
\end{center}

\begin{center}
\vspace{0.2cm}
\quad  \hspace{-0.8cm}  
\infer[\scriptstyle\vee I_{1}]{inl(t^{+})^{+}: A \vee B}{\infer{t^{+}:A}{\infer*{}{(\Gamma;\Delta)}}}
\quad  \quad \quad
\infer[\scriptstyle\vee I_{2}]{inr(t^{+})^{+}:A \vee B}{\infer{t^{+}:B}{\infer*{}{(\Gamma;\Delta)}}}
\end{center}

\begin{center}
\vspace{0.2cm}
\hspace{-0.4cm} 
\deduce[\scriptstyle-----------------------------------\scriptstyle\vee E]{\texttt{case}~ r^{+}\{x^{+}.s^{*} | y^{+}.t^{*}\}^{*}: C}{\infer{r^{+}:A \vee B}{\infer*{}{(\Gamma';\Delta')}} \quad \deduce[\scriptstyle---]{s^{*}:C}{\infer*{}{(\lbrack x^{+}:A\rbrack, \Gamma''; \Delta'')}} \quad \deduce[\scriptstyle---]{t^{*}:C}{\infer*{}{(\lbrack y^{+}: B \rbrack, \Gamma''';\Delta''')}}}
\end{center}

\begin{center}
\vspace{0.2cm}
\quad \hspace{-0.8cm} 
\infer=[\scriptstyle\vee I^{d}]{\langle s^{-}, t^{-} \rangle^{-}:A \vee B}{\;\infer={s^{-}:A}{\infer*{}{(\Gamma';\Delta')}}\quad \quad\infer={t^{-}:B}{\infer*{}{(\Gamma'';\Delta'')}} }
\quad  \quad \quad
\infer=[\scriptstyle\vee E^{d}_{1}]{fst(t^{-})^{-}:A}{\infer={t^{-}:A \vee B}{\infer*{}{(\Gamma;\Delta)}}}
\quad  \quad \quad
\infer=[\scriptstyle\vee E^{d}_{2}]{snd(t^{-})^{-}:B}{\infer={t^{-}:A \vee B}{\infer*{}{(\Gamma;\Delta)}}}
\end{center}

\begin{center}
\vspace{0.2cm}
\quad  \hspace{-0.8cm}
\infer[\scriptstyle\rightarrow I]{(\lambda x^{+}.t^{+})^{+}:A \rightarrow B}{
	\infer{t^{+}:B}{\infer*{}{([x^{+}:A],\Gamma;\Delta)}}}
\quad \quad \quad
\infer[\scriptstyle\rightarrow E]{App(s^{+},t^{+})^{+}:B}{\infer{s^{+}:A \rightarrow B}{\infer*{}{(\Gamma';\Delta')}} \ \quad \infer{t^{+}:A}{\infer*{}{(\Gamma'';\Delta'')}}}
\end{center}

\begin{center}
\vspace{0.2cm}
\quad  \hspace{-0.8cm}
\infer=[\scriptstyle\rightarrow I^{d}]{\{ s^{+},t^{-} \}^{-}: A \rightarrow B}{\;\infer{s^{+}:A}{\infer*{}{(\Gamma';\Delta')}}\quad \infer={t^{-}:B}{\infer*{}{(\Gamma'';\Delta'')}}}
\quad \quad \quad
\infer[\scriptstyle\rightarrow E^{d}_{1}]{\pi_1(t^{-})^{+}:A}{
	\infer={t^{-}:A \rightarrow B}{\infer*{}{(\Gamma;\Delta)}}}
\quad \quad \quad
\infer=[\scriptstyle\rightarrow E^{d}_{2}]{{\pi_2(t^{-})^{-}}:B}{
	\infer={t^{-}:A \rightarrow B}{\infer*{}{(\Gamma;\Delta)}}}
	\end{center}

\begin{center}
\vspace{0.2cm}
\quad \hspace{-0.8cm} 
\infer[\scriptstyle\Yleft I]{\{ t^{+},s^{-} \}^{+}:B \Yleft A}{\;\infer{t^{+}:B}{\infer*{}{(\Gamma';\Delta')}}\quad \quad\infer={s^{-}:A}{\infer*{}{(\Gamma'';\Delta'')}}}
\quad \quad \quad
\infer[\scriptstyle\Yleft E_{1}]{\pi_1(t^{+})^{+}:B}{\infer{t^{+}:B \Yleft A}{\infer*{}{(\Gamma;\Delta)}}}
\quad \quad \quad 
\infer=[\scriptstyle\Yleft E_{2}]{\pi_2(t^{+})^{-}:A}{
	\infer{t^{+}:B \Yleft A}{\infer*{}{(\Gamma;\Delta)}}}
		\end{center}

\begin{center}
\vspace{0.2cm}
\quad \hspace{-0.8cm}
\infer=[\scriptstyle\Yleft I^{d}]{(\lambda x^{-}.t^{-})^{-}:B \Yleft A}{
	\infer={t^{-}:B}{\infer*{}{(\Gamma; \Delta, \llbracket x^{-}:A \rrbracket)}}}
\quad \quad \quad
\infer=[\scriptstyle\Yleft E^{d}]{App(s^{-},t^{-})^{-}:B}{\;\;\;\;\infer={s^{-}:B \Yleft A}{\infer*{}{(\Gamma';\Delta')}} \quad \quad \infer={t^{-}:A}{\infer*{}{(\Gamma'';\Delta'')}}}
\end{center}

\noindent The following lemmata show how terms of a certain form are typed and we need them to prove the Subject Reduction Theorem as well as our Dualization Theorem. 
The terminology and presentation of the lemmata and proofs are to a great extent in the style of \citep{Barendregt1992} and \citep{SU}.

\begin{definition}
 The height of a derivation is the greatest number of successive applications of rules in it, where assumptions have height 0.
\end{definition}

\begin{lemma}[Generation Lemma]

\hspace{0.5cm} 1. Assumptions

1.1 For every $x$, if $(\Gamma; \Delta) \Rightarrow^{+} x^{+}:A$, then $(x^{+}:A) \in \Gamma $.

1.2 For every $x$, if $(\Gamma; \Delta) \Rightarrow^{-} x^{-}:A$, then $(x^{-}:A) \in \Delta $.\\\

2. $\top/\bot$-rules

2.1 If $(\Gamma; \Delta) \Rightarrow^{+} \texttt{top}^{+}:A$, then  $A \equiv \top$.

2.2 If $(\Gamma; \Delta) \Rightarrow^{-} \texttt{bot}^{-}:A$, then  $A \equiv \bot$.

2.3 If $(\Gamma; \Delta) \Rightarrow^{*} abort(t^{+})^{*}:A$, then $(\Gamma; \Delta) \Rightarrow^{+} t^{+}:\bot$.

2.4 If $(\Gamma; \Delta) \Rightarrow^{*} abort(t^{-})^{*}:A$, then $(\Gamma; \Delta) \Rightarrow^{-} t^{-}:\top$.\\\

3. $\rightarrow$-rules

3.1 If $(\Gamma; \Delta) \Rightarrow^{+} (\lambda x^{+}.t^{+})^{+}:C$, then $\exists A, B[(\Gamma, x^{+}:A; \Delta)\Rightarrow^{+} t^{+}:\\\hspace*{0.5cm} B ~\&~ C \equiv A \rightarrow B]$.

3.2 If $(\Gamma, \Gamma'; \Delta, \Delta') \Rightarrow^{+} App(s^{+}, t^{+})^{+}: B$, then $\exists A[(\Gamma; \Delta)\Rightarrow^{+} s^{+}:\\\hspace*{0.5cm}A \rightarrow B ~\&~ (\Gamma'; \Delta')\Rightarrow^{+} t^{+}:A]$.

3.3 If $(\Gamma, \Gamma'; \Delta, \Delta') \Rightarrow^{-} \{ s^{+}, t^{-}\}^{-}: C$, then $\exists A, B[(\Gamma; \Delta)\Rightarrow^{+} s^{+}:A ~\&~\\\hspace*{0.5cm} (\Gamma'; \Delta')\Rightarrow^{-} t^{-}:B ~\&~ C \equiv A \rightarrow B]$.

3.4 If $(\Gamma; \Delta) \Rightarrow^{+} \pi_1(t^{-})^{+}: A$, then $\exists B[(\Gamma; \Delta)\Rightarrow^{-} t^{-}:A \rightarrow B]$.

3.5 If $(\Gamma; \Delta) \Rightarrow^{-} \pi_2(t^{-})^{-}: B$, then $\exists A[(\Gamma; \Delta)\Rightarrow^{-} t^{-}:A \rightarrow B]$.\\\

4. $\Yleft$-rules

4.1 If $(\Gamma, \Gamma'; \Delta, \Delta') \Rightarrow^{+} \{ s^{+}, t^{-}\}^{+}: C$, then $\exists A, B[(\Gamma; \Delta)\Rightarrow^{+} s^{+}:B ~\&~ \\\hspace*{0.5cm}(\Gamma'; \Delta')\Rightarrow^{-} t^{-}:A ~\&~ C \equiv B \Yleft A]$.

4.2  If $(\Gamma; \Delta) \Rightarrow^{+} \pi_1(t^{+})^{+}: B$, then $\exists A[(\Gamma; \Delta)\Rightarrow^{+} t^{+}:B \Yleft A]$.

4.3 If $(\Gamma; \Delta) \Rightarrow^{-} \pi_2(t^{+})^{-}: A$, then $\exists B[(\Gamma; \Delta)\Rightarrow^{+} t^{+}:B \Yleft A]$.

4.4 If $(\Gamma; \Delta) \Rightarrow^{-} (\lambda x^{-}.t^{-})^{-}:C$, then $\exists A, B[(\Gamma; \Delta, x^{-}:A)\Rightarrow^{-}\\\hspace*{0.5cm} t^{-}:B ~\&~ C \equiv B \Yleft A]$.

4.5 If $(\Gamma, \Gamma'; \Delta, \Delta') \Rightarrow^{-} App(s^{-}, t^{-})^{-}: B$, then $\exists A[(\Gamma; \Delta)\Rightarrow^{-} \\\hspace*{0.5cm}s^{-}:B \Yleft A ~\&~ (\Gamma'; \Delta')\Rightarrow^{-} t^{-}:A]$.\\\

5. $\wedge$-rules

5.1 If $(\Gamma, \Gamma'; \Delta, \Delta') \Rightarrow^{+} \langle s^{+}, t^{+}\rangle^{+}: C$, then $\exists A, B[(\Gamma; \Delta)\Rightarrow^{+} s^{+}:A ~\&~ \\\hspace*{0.5cm}(\Gamma'; \Delta')\Rightarrow^{+} t^{+}:B ~\&~ C \equiv A \wedge B]$.

5.2 If $(\Gamma; \Delta) \Rightarrow^{+} fst(t^{+})^{+}: A$, then $\exists B[(\Gamma; \Delta)\Rightarrow^{+} t^{+}:A \wedge B]$.

5.3 If $(\Gamma; \Delta) \Rightarrow^{+} snd(t^{+})^{+}: B$, then $\exists A[(\Gamma; \Delta)\Rightarrow^{+} t^{+}:A \wedge B]$.

5.4 If $(\Gamma; \Delta) \Rightarrow^{-} inl(t^{-})^{-}: C$, then $\exists A, B[(\Gamma; \Delta)\Rightarrow^{-} t^{-}: A ~\&~ C \equiv A \wedge B]$.

5.5 If $(\Gamma; \Delta) \Rightarrow^{-} inr(t^{-})^{-}: C$, then $\exists A, B[(\Gamma; \Delta)\Rightarrow^{-} t^{-}: B ~\&~ C \equiv A \wedge B]$.

5.6 If $(\Gamma, \Gamma', \Gamma''; \Delta, \Delta', \Delta'') \Rightarrow^{*} \texttt{case}~ r^{-}\{x^{-}.s^{*} | y^{-}.t^{*}\}^{*}: C$, then \\\hspace*{0.5cm}$\exists A, B[(\Gamma; \Delta)\Rightarrow^{-} r^{-}: A \wedge B ~\&~ (\Gamma'; \Delta', x^{-}:A)\Rightarrow^{*} s^{*}:C ~\&~ \\\hspace*{0.5cm}(\Gamma''; \Delta'', y^{-}:B)\Rightarrow^{*} t^{*}:C]$.\\\

6. $\vee$-rules

6.1 If $(\Gamma; \Delta) \Rightarrow^{+} inl(t^{+})^{+}: C$, then $\exists A, B[(\Gamma; \Delta)\Rightarrow^{+} t^{+}: A ~\&~ C \equiv A \vee B]$.

6.2 If $(\Gamma; \Delta) \Rightarrow^{+} inr(t^{+})^{+}: C$, then $\exists A, B[(\Gamma; \Delta)\Rightarrow^{+} t^{+}: B ~\&~ C \equiv A \vee B]$.

6.3 If $(\Gamma, \Gamma', \Gamma''; \Delta, \Delta', \Delta'') \Rightarrow^{*} \texttt{case}~ r^{+}\{x^{+}.s^{*} | y^{+}.t^{*}\}^{*}: C$, then \\\hspace*{0.5cm}$\exists A, B[(\Gamma; \Delta)\Rightarrow^{+} r^{+}: A \vee B ~\&~ (\Gamma', x^{+}:A; \Delta')\Rightarrow^{*} s^{*}:C ~\&~ \\\hspace*{0.5cm}(\Gamma'', y^{+}:B; \Delta'')\Rightarrow^{*} t^{*}:C]$.

6.4 If $(\Gamma, \Gamma'; \Delta, \Delta') \Rightarrow^{-} \langle s^{-}, t^{-}\rangle^{-}: C$, then $\exists A, B[(\Gamma; \Delta)\Rightarrow^{-} s^{-}:A ~\&~ \\\hspace*{0.5cm}(\Gamma'; \Delta')\Rightarrow^{-} t^{-}:B ~\&~ C \equiv A \vee B]$.

6.5 If $(\Gamma; \Delta) \Rightarrow^{-} fst(t^{-})^{-}: A$, then $\exists B[(\Gamma; \Delta)\Rightarrow^{-} t^{-}:A \vee B]$.

6.6 If $(\Gamma; \Delta) \Rightarrow^{-} snd(t^{-})^{-}: B$, then $\exists A[(\Gamma; \Delta)\Rightarrow^{-} t^{-}:A \vee B]$.\\\

\begin{proof}
By induction on the height $n$ of the derivation.
If $n=0$, then $(\Gamma; \Delta) \Rightarrow^{*} t^{*}:A$ must consist of either an arbitrary single assumption, in which case $t\equiv x$ and (by definition of the basis $(\Gamma; \Delta)$ in the rules) either $(x^{+}:A) \in \Gamma$ or $(x^{-}:A) \in \Delta$, or $\Gamma=\Delta=\emptyset$ and $t^{*}\equiv$ \texttt{top}$^{+}$ with $A\equiv \top$ or $t^{*}\equiv$ \texttt{bot}$^{-}$ with $A\equiv\bot$.

\noindent Assume now that the clauses of the Generation Lemma hold for all derivations of height $n$.
Then for all clauses 2.3-6.6 the following holds: If there is a derivation of height $n+1$ of the form given on the left side of the implication, then by the rules given above for \texttt{N2Int$_\lambda$} there must be a derivation of height $n$ of the form given on the right side of the implication.

\end{proof}
\end{lemma}

\begin{lemma}[Substitution Lemma]

1. If $(\Gamma, x^{+}:A; \Delta) \Rightarrow^{*} t^{*}: B$ and $(\Gamma'; \Delta') \Rightarrow^{+} s^{+}: A$, then $(\Gamma, \Gamma'; \Delta, \Delta') \Rightarrow^{*} t[s/x]^{*}: B$.\\
2. If $(\Gamma; \Delta, x^{-}:A) \Rightarrow^{*} t^{*}: B$ and $(\Gamma'; \Delta') \Rightarrow^{-} s^{-}: A$, then $(\Gamma, \Gamma'; \Delta, \Delta') \Rightarrow^{*} t[s/x]^{*}: B$.
\begin{proof}
 By induction on the generation of $(\Gamma, x^{+}:A; \Delta) \Rightarrow^{*} t^{*}: B$, respectively $(\Gamma; \Delta, x^{-}:A) \Rightarrow^{*} t^{*}: B$.\\
For the base cases (we'll leave the even more trivial cases, where we have a proof of $\top$ and a refutation of $\bot$, out), we would have\\
$(\Gamma, x^{+}:A; \Delta) \Rightarrow^{+} x^{+}: A$ and $(\Gamma'; \Delta') \Rightarrow^{+} s^{+}: A$, respectively \\
$(\Gamma; \Delta, x^{-}:A) \Rightarrow^{-} x^{-}: A$ and $(\Gamma'; \Delta') \Rightarrow^{-} s^{-}: A$.\\
Thus, trivially (since monotonicity is assumed in the system)\\
$(\Gamma, \Gamma'; \Delta, \Delta') \Rightarrow^{+} s^{+}: A$, respectively\\
$(\Gamma, \Gamma'; \Delta, \Delta') \Rightarrow^{-} s^{-}: A$.\\
We will consider two exemplary cases (choosing ones with a mixture of polarities to make it more interesting) to show that the proof is straightforward. For clause 1, let us consider the case that the last rule applied is $\Yleft E_2$. Then we have\\
$(\Gamma, x^{+}:A; \Delta) \Rightarrow^{-} \pi_2(r^{+})^{-}: B$ and $(\Gamma'; \Delta') \Rightarrow^{+} s^{+}: A$.\\
Thus, by Generation Lemma 4.3, for some $C$\\
$(\Gamma, x^{+}:A; \Delta) \Rightarrow^{+} r^{+}:  C \Yleft B$.\\
By our inductive hypothesis\\
$(\Gamma, \Gamma'; \Delta, \Delta') \Rightarrow^{+} r[s/x]^{+}: C \Yleft B$.\\
Thus, by $\Yleft E_2$\\
$(\Gamma, \Gamma'; \Delta, \Delta') \Rightarrow^{-} \pi_2(r[s/x]^{+})^{-}: B$. \\
For clause 2, let us consider the case that the last rule applied is $\wedge I$. Then we have
$(\Gamma; \Delta, x^{-}:A) \Rightarrow^{+} \langle r^{+}, u^{+}\rangle^{+}: C \wedge D$ and $(\Gamma'; \Delta') \Rightarrow^{-} s^{-}: A$.\\
Thus, by Generation Lemma 5.1 with $\Gamma = \Gamma'' \cup \Gamma'''$ and $\Delta = \Delta'' \cup \Delta'''$ either\\
$(\Gamma''; \Delta'', x^{-}:A) \Rightarrow^{+} r^{+}: C $ and $(\Gamma'''; \Delta''') \Rightarrow^{+} u^{+}: D $ or\\
$(\Gamma''; \Delta'') \Rightarrow^{+} r^{+}: C $ and $(\Gamma'''; \Delta''', x^{-}:A) \Rightarrow^{+} u^{+}: D $.\\
By our inductive hypothesis then either\\
$(\Gamma', \Gamma''; \Delta', \Delta'') \Rightarrow^{+} r[s/x]^{+}: C $ or\\
$(\Gamma', \Gamma'''; \Delta', \Delta''') \Rightarrow^{+} u[s/x]^{+}: D $.\\
Thus, by $\wedge I$ either\\
$(\Gamma, \Gamma'; \Delta, \Delta')\Rightarrow^{+} \langle r[s/x]^{+}, u^{+}\rangle^{+}: C \wedge D$ or\\
$(\Gamma, \Gamma'; \Delta, \Delta') \Rightarrow^{+} \langle r^{+}, u[s/x]^{+}\rangle^{+}: C \wedge D$.
\end{proof}
\end{lemma}

\noindent Let us now consider the reductions available in our framework.
For their definition the inductive definition of a \emph{compatible} relation will be outsourced to the appendix because for $\lambda^{2Int}$ we need a lot of clauses (cf. Definition \ref{Compatibility}).
Suffice it to say that a ``compatible relation `respects' the syntactic constructions'' \cite[p. 12]{SU} of the terms, i.e., let $\mathcal{R}$ be a compatible relation on \texttt{Term$_{2Int}$}, then for all $t, r, s \in$ \texttt{Term$_{2Int}$}: if $t \mathcal{R} r$, then $(\lambda x^{*}.t^{*})^{*} \mathcal{R} (\lambda x^{*}.r^{*})^{*}$, $App(t^{*}, s^{*})^{*} \mathcal{R} App(r^{*}, s^{*})^{*}$, $App(s^{*}, t^{*})^{*} \mathcal{R} App(s^{*}, r^{*})^{*}$, etc.

\begin{definition}[Reductions]

1. The least compatible relation $\rightsquigarrow_{1\beta}$ on \texttt{Term$_{2Int}$} satisfying the following clauses is called \emph{$\beta$-reduction}:

$App((\lambda x^{*}.t^{*})^{*}, s^{*})^{*}$ $\rightsquigarrow_{1\beta}$ $t[s^{*}/x^{*}]^{*}$

$\pi_1(\{ s^{+},t^{-}\}^{*})^{+}$ $\rightsquigarrow_{1\beta}$ $s^{+}$
\quad
$\pi_2(\{s^{+},t^{-}\}^{*})^{-}$ $\rightsquigarrow_{1\beta}$ $t^{-}$

$fst(\langle s^{*},t^{*}\rangle^{*})^{*}$ $\rightsquigarrow_{1\beta}$ $s^{*}$
\quad 
$snd(\langle s^{*},t^{*}\rangle^{*})^{*}$ $\rightsquigarrow_{1\beta}$ $t^{*}$

$\texttt{case}~ inl(r^{*})^{*} \{x^{*}.s^{\dagger} | y^{*}.t^{\dagger}\}^{\dagger}$ $\rightsquigarrow_{1\beta}$ $s[r^{*}/x^{*}]^{\dagger}$

$\texttt{case}~ inr(r^{*})^{*} \{x^{*}.s^{\dagger} | y^{*}.t^{\dagger}\}^{\dagger}$ $\rightsquigarrow_{1\beta}$ $t[r^{*}/y^{*}]^{\dagger}$\\
2. For all clauses the term on the left of $\rightsquigarrow_{1\beta}$ is called \emph{$\beta$-redex}, while the term on the right is its \emph{contractum}.\\
3. The relation $\rightsquigarrow_\beta$ (multi-step $\beta$-reduction) is the transitive and reflexive closure of $\rightsquigarrow_{1\beta}$.
\end{definition}

\noindent Philosophically important (cf. Section 3) is that the reduction relation is type-preserving, which is established with the following subject reduction theorem for this calculus.

\begin{theorem}[Subject Reduction Theorem for $\lambda^{2Int}$]
If $(\Gamma; \Delta) \Rightarrow^{*} t^{*}: C$ and $t \rightsquigarrow_\beta t'$, then $(\Gamma'; \Delta') \Rightarrow^{*} t'^{*}: C$ for $\Gamma' \subseteq \Gamma$ and $\Delta' \subseteq \Delta$.
\begin{proof}
By induction on the generation of $\rightsquigarrow_\beta$ using the Generation and Substitution Lemmata. \\
We will spell out one of the reductions for each connective. For the connectives where we have two reductions it will be straightforward that the same reasoning can be applied.\\
Suppose $t \equiv App((\lambda x^{*}.r^{*})^{*}, s^{*})^{*}$, $t' \equiv r[s^{*}/x^{*}]^{*}$, $\Gamma = \Gamma' \cup \Gamma''$ and $\Delta = \Delta' \cup \Delta''$.
If
$(\Gamma; \Delta) \Rightarrow^{*} App((\lambda x^{*}.r^{*})^{*}, s^{*})^{*}: C$,\\
then by Generation Lemma 3.2 and 4.5 there must be some $A$ such that either\\ 
$(\Gamma'; \Delta')\Rightarrow^{+} (\lambda x^{+}.r^{+})^{+}:A \rightarrow C$ and $(\Gamma''; \Delta'')\Rightarrow^{+} s^{+}:A$, or\\
$(\Gamma'; \Delta')\Rightarrow^{-} (\lambda x^{-}.r^{-})^{-}:C \Yleft A$ and $(\Gamma''; \Delta'')\Rightarrow^{-} s^{-}:A$.\\
Thus, again by Generation Lemma 3.1 and 4.4 it follows that either \\
$(\Gamma',x^{+}:A; \Delta')\Rightarrow^{+} r^{+}: C$ and $(\Gamma''; \Delta'')\Rightarrow^{+} s^{+}:A$, or\\
$(\Gamma'; \Delta',x^{-}:A)\Rightarrow^{-} r^{-}: C$ and $(\Gamma''; \Delta'')\Rightarrow^{-} s^{-}:A$.\\
Therefore, by the Substitution Lemma either\\
$(\Gamma; \Delta) \Rightarrow^{+} r[s^{+}/x^{+}]^{+}: C$, or\\
$(\Gamma; \Delta) \Rightarrow^{-} r[s^{-}/x^{-}]^{-}: C$.
\vspace{0.1cm}

\noindent Suppose $t \equiv \pi_1(\{ s^{+},r^{-}\}^{*})^{+}$, $t' \equiv s^{+}$, $\Gamma' \subseteq \Gamma$ and $\Delta' \subseteq \Delta$.\\
If $(\Gamma; \Delta) \Rightarrow^{+} \pi_1(\{ s^{+},r^{-}\}^{*})^{+}: C$,\\
then by Generation Lemma 3.4 and 4.2 there must be some $A$ such that either\\ 
$(\Gamma; \Delta) \Rightarrow^{-} \{ s^{+},r^{-}\}^{-}: C \rightarrow A$, or\\
$(\Gamma; \Delta) \Rightarrow^{+} \{ s^{+},r^{-}\}^{+}: C \Yleft A$.\\
Thus, again by Generation Lemma 3.3 and 4.1 it follows that in both cases \\
$(\Gamma'; \Delta')\Rightarrow^{+} s^{+}: C$.
\vspace{0.1cm}

\noindent Suppose $t \equiv fst(\langle s^{*},r^{*}\rangle^{*})^{*}$, $t' \equiv s^{*}$, $\Gamma' \subseteq \Gamma$ and $\Delta' \subseteq \Delta$.\\
If $(\Gamma; \Delta) \Rightarrow^{*} fst(\langle s^{*},r^{*}\rangle^{*})^{*}: C$,\\
then by Generation Lemma 5.2 and 6.5 there must be some $A$ such that either\\ 
$(\Gamma; \Delta) \Rightarrow^{+} \langle s^{+},r^{+}\rangle^{+}: C \wedge A$, or\\
$(\Gamma; \Delta) \Rightarrow^{-} \langle s^{-},r^{-}\rangle^{-}: C \vee A$.\\
Thus, again by Generation Lemma 5.1 and 6.4 it follows that in both cases \\
$(\Gamma'; \Delta')\Rightarrow^{*} s^{*}: C$.
\vspace{0.1cm}

\noindent Suppose $t \equiv \texttt{case}~ inl(r^{*})^{*} \{x^{*}.s^{\dagger} | y^{*}.u^{\dagger}\}^{\dagger}$, $t' \equiv s[r^{*}/x^{*}]^{\dagger}$, $\Gamma = \Gamma' \cup \Gamma''\cup \Gamma'''$ and $\Delta = \Delta' \cup \Delta''\cup \Delta'''$.\\
If $(\Gamma; \Delta) \Rightarrow^{\dagger} \texttt{case}~ inl(r^{*})^{*} \{x^{*}.s^{\dagger} | y^{*}.u^{\dagger}\}^{\dagger}: C$,\\
then by Generation Lemma 5.6 and 6.3 there must be some $A$ and $B$ such that either\\
$(\Gamma'; \Delta')\Rightarrow^{+} inl(r^{+})^{+}: A \vee B$ and $(\Gamma'', x^{+}:A; \Delta'')\Rightarrow^{\dagger} s^{\dagger}:C$ and $(\Gamma''', y^{+}:B; \Delta''')\Rightarrow^{\dagger} u^{\dagger}:C$, or\\
$(\Gamma'; \Delta')\Rightarrow^{-} inl(r^{-})^{-}: A \wedge B$ and $(\Gamma''; \Delta'', x^{-}:A)\Rightarrow^{\dagger} s^{\dagger}:C$ and $(\Gamma'''; \Delta''', y^{-}:B)\Rightarrow^{\dagger} u^{\dagger}:C$.\\
Thus, again by Generation Lemma 5.4 and 6.1 it follows that either \\
$(\Gamma'; \Delta')\Rightarrow^{+} r^{+}: A$, or\\
$(\Gamma'; \Delta')\Rightarrow^{-} r^{-}: A$.\\
Therefore, by the Substitution Lemma either\\
$(\Gamma', \Gamma''; \Delta', \Delta'')\Rightarrow^{\dagger} s[r^{+}/x^{+}]^{\dagger}: C$, or\\
$(\Gamma', \Gamma''; \Delta', \Delta'')\Rightarrow^{\dagger} s[r^{-}/x^{-}]^{\dagger}: C$.
\end{proof}

\end{theorem}

\noindent Due to the structure of the disjunction elimination rule and the dual conjunction elimination rule, we also need so-called \emph{permutation} and \emph{simplification conversions} next to $\beta$-reductions:

\begin{definition}[Permutation conversions]

1. The least compatible relation $\rightsquigarrow_{1 p}$ on \texttt{Term$_{2Int}$} satisfying the following clauses is called \emph{permutation conversion}:

$App(\texttt{case}~ r^{*} \{x^{*}.s^{\dagger} | y^{*}.t^{\dagger}\}^{\dagger}, u^{\dagger})^{\dagger}$ $\rightsquigarrow_{1 p}$ $\texttt{case}~ r^{*} \{x^{*}.App(s^{\dagger},u^{\dagger})^{\dagger} | y^{*}.App(t^{\dagger}, u^{\dagger})^{\dagger}\}^{\dagger}$

$\pi_1(\texttt{case}~ r^{*} \{x^{*}.s^{\dagger} | y^{*}.t^{\dagger}\}^{\dagger})^{+}$ $\rightsquigarrow_{1 p}$ $\texttt{case}~ r^{*} \{x^{*}.\pi_1(s^{\dagger})^{+} | y^{*}.\pi_1(t^{\dagger})^{+}\}^{+}$

$\pi_2(\texttt{case}~ r^{*} \{x^{*}.s^{\dagger} | y^{*}.t^{\dagger}\}^{\dagger})^{-}$ $\rightsquigarrow_{1 p}$ $\texttt{case}~ r^{*} \{x^{*}.\pi_2(s^{\dagger})^{-} | y^{*}.\pi_2(t^{\dagger})^{-}\}^{-}$

$fst(\texttt{case}~ r^{*} \{x^{*}.s^{\dagger} | y^{*}.t^{\dagger}\}^{\dagger})^{\dagger}$ $\rightsquigarrow_{1 p}$ $\texttt{case}~ r^{*} \{x^{*}.fst(s^{\dagger})^{\dagger} | y^{*}.fst(t^{\dagger})^{\dagger}\}^{\dagger}$
 
$snd(\texttt{case}~ r^{*} \{x^{*}.s^{\dagger} | y^{*}.t^{\dagger}\}^{\dagger})^{\dagger}$ $\rightsquigarrow_{1 p}$ $\texttt{case}~ r^{*} \{x^{*}.snd(s^{\dagger})^{\dagger} | y^{*}.snd(t^{\dagger})^{\dagger}\}^{\dagger}$

$\texttt{case}~ \texttt{case}~ r^{*} \{x^{*}.s^{\dagger} | y^{*}.t^{\dagger}\}^{\dagger}\{z_1^{\dagger}.u^{+} | z_2^{\dagger}.v^{+}\}^{+}$ $\rightsquigarrow_{1 p}$

$\texttt{case}~ r^{*} \{x^{*}.\texttt{case}~ s^{\dagger}\{z_1^{\dagger}.u^{+} | z_2^{\dagger}.v^{+}\}^{+} | y^{*}.\texttt{case}~ t^{\dagger}\{z_1^{\dagger}.u^{+} | z_2^{\dagger}.v^{+}\}^{+}\}^{+}$

$\texttt{case}~ \texttt{case}~ r^{*} \{x^{*}.s^{\dagger} | y^{*}.t^{\dagger}\}^{\dagger}\{z_1^{\dagger}.u^{-} | z_2^{\dagger}.v^{-}\}^{-}$ $\rightsquigarrow_{1 p}$

$\texttt{case}~ r^{*} \{x^{*}.\texttt{case}~ s^{\dagger}\{z_1^{\dagger}.u^{-} | z_2^{\dagger}.v^{-}\}^{-} | y^{*}.\texttt{case}~ t^{\dagger}\{z_1^{\dagger}.u^{-} | z_2^{\dagger}.v^{-}\}^{-}\}^{-}$
\\
2. For all clauses the term on the left of $\rightsquigarrow_{1 p}$ is called \emph{p-redex}.\\
3. The relation $\rightsquigarrow_p$ (multi-step permutation conversion) is the transitive and reflexive closure of $\rightsquigarrow_{1 p}$.
\end{definition}

\begin{definition}[Simplification conversion]

1. The least compatible relation $\rightsquigarrow_{1 s}$ on \texttt{Term$_{2Int}$} satisfying the following clause is called \emph{s-conversion}:

$\texttt{case}~ r^{*} \{x^{*}.s_1^{\dagger} | y^{*}.s_2^{\dagger}\}^{\dagger}$ $\rightsquigarrow_{1 s}$ $s_i^{\dagger}$ $(i \in \{1, 2\})$ where $x^{*}$ and $y^{*}$ do not occur freely in $s_i^{\dagger}$.

\noindent 2. The term on the left of $\rightsquigarrow_{1 s}$ is called \emph{s-redex}.\\
3. The relation $\rightsquigarrow_s$ (multi-step simplification conversion) is the transitive and reflexive closure of $\rightsquigarrow_{1 s}$.
\end{definition}

\noindent As with $\beta$-reductions, both permutation and simplification conversions  satisfy the property of type-preservation.

\begin{definition}[Normal form]

A term $t \in$ \texttt{Term$_{2Int}$} is said to be in \emph{normal form} iff $t$ does not contain any $\beta$-, p-, or s-redex.

\end{definition}

\noindent From now on we will omit the superscripts of subterms in the cases where the superscript of the whole term clearly determines the other polarities, i.e., instead of, e.g., $(\lambda x^{+}.t^{+})^{+}$ writing $(\lambda x.t)^{+}$ suffices.

\subsection{Duality in $\lambda^{2Int}$}
I want to examine now a bit closer how the polarities in $\lambda^{2Int}$ relate to each other, and thereby, more generally speaking, the relation between proofs and refutations in this system.
Therefore, I will define dualities in $\lambda^{2Int}$ and then prove our Dualization Theorem, which will be the key feature for the philosophical implications discussed in the next section.
\begin{definition}[Duality]\label{duality}
We will define a duality function $d$ mapping types to their dual types, terms to their dual terms and bases to their dual bases as follows:\footnote{The superscript $^{d}$ is used to indicate the dual polarity of whatever polarity $^{*}$ stands for in its respective dual version.}

1. $d(\rho) = \rho$

2.  $d(\top) = \bot$

3. $d(\bot) = \top$

4. $d(A \wedge B) = d(A) \vee d(B)$ 

5. $d(A \vee B) = d(A) \wedge d(B)$ 

6. $d(A \rightarrow B) = d(B) \Yleft d(A)$ 

7.  $d(A \Yleft B) = d(B) \rightarrow d(A)$ 

8. $d(x^{*}) = x^{d}$

9. $d(\texttt{top}^{+}) = \texttt{bot}^{-}$

10. $d(\texttt{bot}^{-}) = \texttt{top}^{+}$

11. $d(abort(t^{*})^{\dagger}) = abort(d(t^{*}))^{d}$

12. $d(\langle t^{*},s^{*} \rangle^{*}) = \langle d(t^{*}),d(s^{*}) \rangle^{d}$

13. $d(inl(t^{*})^{*}) = inl(d(t^{*}))^{d}$

14. $d(inr(t^{*})^{*}) = inr(d(t^{*}))^{d}$

15. $d((\lambda x^{*}.t^{*})^{*}) = (\lambda d(x^{*}).d(t^{*}))^{d}$

16. $d(\{ t^{+},s^{-} \}^{*}) = \{ d(s^{-}),d(t^{+}) \}^{d}$

17. $d(fst(t^{*})^{*}) = fst(d(t^{*}))^{d}$

18. $d(snd(t^{*})^{*}) = snd(d(t^{*}))^{d}$

19. $d(\texttt{case}~ r^{*}\{x^{*}.s^{\dagger} | y^{*}.t^{\dagger}\}^{\dagger}) = \texttt{case}~ d(r^{*})\{d(x^{*}).d(s^{\dagger}) | d(y^{*}).d(t^{\dagger})\}
^{d}$

20. $d(App(s^{*},t^{*})^{*}) = App(d(s^{*}),d(t^{*}))^{d}$

21. $d(\pi_1(t^{*})^{\dagger}) = \pi_2(d(t^{*}))^{d}$

22. $d(\pi_2(t^{*})^{\dagger}) = \pi_1(d(t^{*}))^{d}$

23. $d((\Gamma; \Delta)) = (d(\Delta); d(\Gamma))$,with $d(\Delta) = \{d(t^{*}) \mid t^{*} \in \Delta\}$, resp. for $d(\Gamma)$.
\end{definition}

\begin{theorem}[Dualization]
If $(\Gamma; \Delta) \Rightarrow^{*} t^{*}:A$ with a height of derivation at most $n$, then $(d(\Delta); d(\Gamma))\Rightarrow^{d} d(t^{*}): d(A)$ (called its \emph{dual derivation}) with a height of derivation at most $n$. This means that whenever we have a proof (refutation) of a formula, we can construct a refutation (proof) with the same height of its dual formula in our system.

\begin{proof}[Proof]\label{duallemma}
By induction on the height of derivation $n$ using the Generation Lemma.

\noindent If $n=0$, then one of the four cases holds:
\begin{enumerate}
\item $(x^{+}: A; \emptyset) \Rightarrow^{+} x^{+}:A$
\item $(\emptyset; x^{-}:A) \Rightarrow^{-} x^{-}:A$
\item $(\emptyset; \emptyset) \Rightarrow^{+} \texttt{top}^{+}:\top$
\item $(\emptyset; \emptyset) \Rightarrow^{-} \texttt{bot}^{-}:\bot$
\end{enumerate}

\noindent In case 1 the dual derivation is $(\emptyset; x^{-}:d(A)) \Rightarrow^{-} x^{-}:d(A)$.\\
\noindent In case 2 the dual derivation is $(x^{+}: d(A); \emptyset) \Rightarrow^{+} x^{+}:d(A)$.\\
In case 3 the dual derivation is $(\emptyset; \emptyset) \Rightarrow^{-} \texttt{bot}^{-}:\bot$.\\
In case 4 the dual derivation is $(\emptyset; \emptyset) \Rightarrow^{+} \texttt{top}^{+}:\top$.\\
All dual derivations can be trivially constructed with a height of $n=0$.

\noindent Assume height-preserving dualization up to derivations of height at most $n$.\\
If $(\Gamma; \Delta) \Rightarrow^{*} abort(t^{+})^{*}:\rho$ is of height $n+1$, then (by Generation Lemma 2.3) we have $(\Gamma; \Delta) \Rightarrow^{+} t^{+}:\bot$ with height at most $n$. 
If $(\Gamma; \Delta) \Rightarrow^{*} abort(t^{-}))^{*}:\rho$ is of height $n+1$, then (by Generation Lemma 2.4) we have $(\Gamma; \Delta) \Rightarrow^{-} t^{-}:\top$ with height at most $n$. 

\noindent Then, by inductive hypothesis $(d(\Delta); d(\Gamma))\Rightarrow^{-} d(t^{+}): d(\bot)$, resp. \\$(d(\Delta); d(\Gamma))\Rightarrow^{+} d(t^{-}): d(\top)$ are of height at most $n$ as well.
  
\noindent By application of $\top E^{d}$, resp. $\bot E$, we can construct a derivation of height $n+1$ s.t. $(d(\Delta); d(\Gamma))\Rightarrow^{*} abort(d(t^{+}))^{*}: \rho$, resp. $(d(\Delta); d(\Gamma))\Rightarrow^{*} abort(d(t^{-}))^{*}: \rho$ with $^*$ being the dual polarity of $^*$ in the original derivations. 
By our definition of dual terms $d(abort(t^{+})^{*}) = abort(d(t^{+}))^{d}$ and $d(abort(t^{-})^{*}) = abort(d(t^{-}))^{d}$.

\vspace{0.1cm}

\noindent If $(\Gamma; \Delta) \Rightarrow^{+} \langle s^{+}, t^{+}\rangle^{+}: A \wedge B$, resp. $(\Gamma; \Delta) \Rightarrow^{-} \langle s^{-}, t^{-}\rangle^{-}: A \vee B$ is of height $n+1$, then (by Generation Lemma 5.1, resp. 6.4) for $\Gamma = \Gamma' \cup \Gamma''$ and $\Delta = \Delta' \cup \Delta''$ we have $(\Gamma'; \Delta')\Rightarrow^{+} s^{+}:A$ and $(\Gamma''; \Delta'')\Rightarrow^{+} t^{+}:B$, resp. $(\Gamma'; \Delta')\Rightarrow^{-} s^{-}:A$ and $(\Gamma''; \Delta'')\Rightarrow^{-} t^{-}:B$ with height at most $n$. 

\noindent Then, by inductive hypothesis $(d(\Delta'); d(\Gamma'))\Rightarrow^{-} d(s^{+}): d(A)$ and \\$(d(\Delta''); d(\Gamma''))\Rightarrow^{-} d(t^{+}): d(B)$, resp. $(d(\Delta'); d(\Gamma'))\Rightarrow^{+} d(s^{-}): d(A)$ and $(d(\Delta''); d(\Gamma''))\Rightarrow^{+} d(t^{-}): d(B)$ are of height at most $n$ as well.
  
\noindent By application of $\vee I^{d}$, resp. $\wedge I$, we can construct a derivation of height $n+1$ s.t. $(d(\Delta); d(\Gamma))\Rightarrow^{-} \langle d(s^{+}), d(t^{+})\rangle^{-}: d(A) \vee d(B)$, resp. $(d(\Delta); d(\Gamma))\Rightarrow^{+} \langle d(s^{-}), d(t^{-})\rangle^{+}: d(A) \wedge d(B)$. 
By our definition of dual terms $d(\langle s^{*}, t^{*}\rangle^{*}) = \langle d(s^{*}), d(t^{*})\rangle^{d}$.\\
For the further cases, see Appendix.
\end{proof}
\end{theorem}

\noindent Let us have a look at an example here considering the following derivation:
\vspace{0.2cm}

\quad  
\infer[\scriptstyle\rightarrow I]{(\lambda x^{+}.\{ \texttt{top}^{+},\{ \pi_1 (x^{+})^{+}, \pi_2(x^{+})^{-} \}^{-}\}^{+})^{+}: (A \Yleft B) \rightarrow (\top \Yleft (A \rightarrow B))} 
{\infer[\scriptstyle\Yleft I]{\{ \texttt{top}^{+}, \{ \pi_1(x^{+})^{+}, \pi_2(x^{+})^{-} \}^{-} \}^{+}: \top \Yleft (A \rightarrow B)}{\;\;\;\;\;\;\;\;\;\infer[\scriptstyle\top I]{\texttt{top}^{+}: \top}{} \quad \infer=[\scriptstyle\rightarrow I^{d}]{\{ \pi_1(x^{+})^{+}, \pi_2(x^{+})^{-} \}^{-} :A \rightarrow B}{\;\;\infer[\scriptstyle\Yleft E_{1}]{\pi_1(x^{+})^{+}: A }{[x^{+}: A \Yleft B]} \quad \quad {\infer=[\scriptstyle\Yleft E_{2}]{\pi_2(x^{+})^{-}:B}{[x^{+}: A \Yleft B]}}}}}
\vspace{0.2cm}

\noindent Now we dualize the term and the formula by our duality function $d$ yielding the following:

\vspace{0.2cm}

$d((\lambda x^{+}.\{ \texttt{top}^{+},\{ \pi_1(x^{+})^{+}, \pi_2(x^{+})^{-} \}^{-}\}^{+})^{+})$ =\\
$(\lambda x^{-}.\{\{  \pi_1(x^{-})^{+}, \pi_2(x^{-})^{-}\}^{+}, \texttt{bot}^{-}\}^{-})^{-}$ 

\vspace{0.2cm}
$d((A \Yleft B) \rightarrow (\top \Yleft (A \rightarrow B))) = ((B \Yleft A) \rightarrow \bot) \Yleft (B \rightarrow A)$

\vspace{0.2cm}

\noindent We can now build a derivation with the dualized term \\$(\lambda x^{-}.\{\{  \pi_1(x^{-})^{+}, \pi_2(x^{-})^{-}\}^{+}, \texttt{bot}^{-}\}^{-})^{-}$ as end-term (i.e., the term annotating the conclusion of a derivation) being of the type of the dualized formula $((B \Yleft A) \rightarrow \bot) \Yleft (B \rightarrow A)$:
\vspace{0.3cm}

\quad
\infer=[\scriptstyle\Yleft I^{d}]{(\lambda x^{-}.\{ \{ \pi_1(x^{-})^{+}, \pi_2(x^{-})^{-} \}^{+}, \texttt{bot}^{-} \}^{-})^{-}: ((B \Yleft A) \rightarrow \bot) \Yleft (B \rightarrow A)} 
{\infer=[\scriptstyle\rightarrow I^{d}]{\{ \{ \pi_1(x^{-})^{+}, \pi_2(x^{-})^{-} \}^{+}, \texttt{bot}^{-} \}^{-}: (B \Yleft A) \rightarrow \bot}{
\infer[\scriptstyle\Yleft I]{\{ \pi_1(x^{-})^{+}, \pi_2(x^{-})^{-} \}^{+} :B \Yleft A}{\;\;{\infer[\scriptstyle\rightarrow E^{d}_{1}]{\pi_1(x^{-})^{+}:B}{\llbracket x^{-}: B \rightarrow A\rrbracket}} \infer=[\scriptstyle\rightarrow E^{d}_{2}]{\pi_2(x^{-})^{-}: A }{\llbracket x^{-}: B \rightarrow A\rrbracket} \quad \quad }\;\;\;\;\;\;\;\;\;\infer=[\scriptstyle\bot I^{d}]{\texttt{bot}^{-}: \bot}{}}}

\vspace{0.2cm}

\vspace{0.2cm}

\noindent The duality between those derivations is literally `visible' in that they look like the mirrored version of each other with respect to the construction of the proof tree and the use of single and double lines.
At each step we have the dual terms with the dual types applied according to the respectively dual rules.
So the case can be made - and this is what I want to argue for in the next section - that what we have in these cases are essentially the \emph{same} underlying constructions, although in one case it is delivered as a proof and in the other as a refutation.

\section{Meaning and identity of derivations in \texttt{N2Int}$_\lambda$ }
\subsection{Background: Distinguishing sense and denotation of proofs}
I will lay out a conception to distinguish in a Fregean way between sense and denotation of proofs based on \citep{Tranchini2016} and \citep{Ayhan2021b}.
The background of this conception is in the area of what has been called \emph{general proof theory}, a program which was strongly advocated in the early 1970s by representatives such as Prawitz \citeyearpar{Prawitz1971}, Kreisel \citeyearpar{Kreisel1971}, and Martin-Löf \citeyearpar{Martin1975}. Their motivation was - in opposition to Hilbert's understanding of proof theory \citep[see][p. 225]{Prawitz1973} - to generate a new appreciation of proofs, namely not as mere tools, which are only studied to achieve specific aims, like establishing the consistency of mathematics, but as objects which are worth to be studied in their own right. 
On such a view it is, e.g., not only an interesting question \emph{what} can be proved, but rather \emph{how} something can be proved. 
In this spirit it is also much more interesting, then, to focus on proof systems containing rules for the logical connectives instead of axiomatic proof systems, which is why their focus shifted toward systems in the style of Gentzen's natural deduction and sequent calculus systems.
These ideas are the basis of proof-theoretic semantics, which can be seen as a specific thread of this program.

\noindent The very general thought then, underlying a (broadly) Fregean distinction of sense and denotation of proofs, is simply that there are different ways to deliver a derivation of the same proof.
A standard example for this would be two derivations, one being in non-normal form and the other being in its respective normal form.
I will distinguish (as it is also done in the literature, e.g., in \cite[p. 110]{Kreisel1971}, \cite[p. 237]{Prawitz1971}, \citep[p. 93]{Martin1975}) for these purposes between a \emph{proof} as the underlying object (conceived of as a mental entity in line of the intuitionistic tradition) and a \emph{derivation} as its respective linguistic representation.
Since the derivation in normal form is the most direct way of representing the proof, it can be argued that this captures the denotation best.
Thus, in the case with a derivation in normal form and one in non-normal form, though reducible to the former, the denotation would be the same since they share the same normal form \citep[p. 257ff.]{Prawitz1971}.
Their sense would differ, however, because the way of representing the denotation is essentially different in these cases.
Such a conception for proofs can be found, e.g., in \cite{Girard1989}, but it is not very developed in the standard literature on this topic.
\cite{Tranchini2016} spelled out in more detail, then, how such a distinction could be usefully applied in the context of PTS.
He argues that a derivation can only have sense if all its rules applied in it have reductions available (as opposed to, e.g., rules for \texttt{tonk}), since the reductions are what transfers a derivation into its normal form, i.e., its denotation.\footnote{Tranchini uses this framework to distinguish between derivations which have sense \emph{and} denotation (`normal', well-behaved proofs), derivations which have sense, yet lack denotation (paradoxical derivations, since reductions can be applied to them but they cannot be brought into normal form by this) and derivations having neither sense nor denotation (which would be, e.g., ones containing a connective like \texttt{tonk}, for which there are no reductions available at all).}
Thus, the reductions are the way to get to the denotation of proofs which seems to fit nicely a Fregean conception of sense.

\noindent \citep{Ayhan2021b} builds upon Tranchini's idea in that the criterion for a derivation to have sense is adopted but also further developed by firstly, transferring it to a setting with $\lambda$-term-annotated proof systems and secondly, giving a concrete account of what constitutes the sense of a derivation.
While the denotation of a derivation in such systems is referred to by the end-term of the derivation and sameness of denotation holds modulo belonging to the same equivalence class induced by the set of conversions\footnote{By this set I mean in our system not only the permutation and simplification conversions but also the $\beta$-reductions, although I do not use the term `conversion' for those. The reason is simply that for the latter the use of `reductions' is more common.} considered for the system, the sense of the derivation is taken to be the set of all $\lambda$-terms occurring within the derivation since these reflect the operations used in the derivation.
Thus, they can be seen as encoding a procedure that takes us to the denotation, since the procedure finally yields the end-term.\footnote{For such an interpretation of Fregean sense, see, e.g., Dummett \citeyearpar[pp. 232, 323, 636]{Dummett1973} speaking of a ``procedure'' to determine the denotation, \citeyearpar[p. 96]{Dummett1973} ``names with different senses but the same referent correspond to different routes leading to the same destination'', Girard \citeyearpar[p. 2]{Girard1989} ``a sequence of \emph{instructions}'' or Horty \citeyearpar[pp. 66-69]{Horty2007} ``senses as procedures''. Girard even mentions this in the context of relating this to understanding ``proofs as programs", i.e., a Curry-Howard conception.}
The benefit of this is that we can thereby not only speak of sameness when it comes to denotation, i.e., identity of proofs, but also about sameness of derivations when it comes to sense, which would be the question of \emph{synonymy}.
Also, by using the $\lambda$-terms we can compare sense and denotation over different kinds of proofs systems, e.g., between natural deduction and sequent calculus systems.
\subsection{Identifying proofs and refutations}

So, what can we say on this basis about sense and denotation of proofs and dual proofs in \texttt{N2Int}$_\lambda$?
 Having established that we have well-behaved reductions (by the Subject Reduction Theorem), we can safely assume that our derivations in \texttt{N2Int}$_\lambda$ do have sense and (although we have not proven a normalization theorem for the terms here) since a normal form theorem is proven for the non-annotated natural deduction system \texttt{N2Int},\footnote{What is more, in \citep{AyhanWansing2023} also the stronger result in form of a cut-elimination theorem is proven for the corresponding sequent calculus \texttt{SC2Int}.} we can assume that they have denotation as well.\footnote{Since we have not proven the confluence property here either, it is not certain that for every derviation there will be a unique term in normal form which is why identity is formulated as being induced modulo belonging to the same equivalence class of the conversions. I would be rather optimistic that the way \texttt{2Int} extends intuitionistic logic does not mess up confluence but the proof would extend the scope of this paper.} 
What I want to argue for is that in this system it seems reasonable to extend our criterion for identity of proofs along the following lines.
What the Dualization Theorem states is that for every proof, resp. dual proof, of a formula, we can express it as one or the other.
This gives us a - from a bilateralist viewpoint - perfect balance in our system: There is no priority for proofs!
Thus, proofs and dual proofs should be viewed as two sides of one coin.
Taking this image seriously, what this amounts to is taking them as different representations of the \emph{same} object, i.e., proofs and refutations of the respective dual formulas are \emph{essentially the same}.
So, my claim is - in Fregean terminology - that those derivations have the same denotation, because the underlying construction is \emph{one} object, but they differ in sense because the way they are represented is essentially different.

\noindent Is it intuitive, though, to identify proofs and refutations, given that they are seemingly rather quite the opposite?
In the traditional literature on falsification, e.g., \cite{Nelson} and \cite{Lopez}, such a thought is indeed expressed, namely that one and the same entity can act as verifying one formula, while falsifying another.
And I also think there are cases from mathematical reasoning or from our empirical way of `proving' something, e.g. in court, where it makes a lot of sense to do so, i.e., where we have one and the same construction/evidence/etc. yielding a proof of a proposition while simultaneously refuting the dual proposition.
If we think of what a proof of the statement ``11 is a prime number'' would look like, it would probably be something along the following lines: A constructive way of showing that 11 is a natural number greater than 1 and is not a product of two smaller natural numbers.
So, basically we could have a program running through all the natural numbers up to 11 and checking whether they could form 11 as a product.\footnote{To clarify a point raised by an anonymous referee: This is just to give one example where I think such an analogy with a program works, it is not meant to suggest that this would work for any program checking any mathematical property for numbers.} 
If this is not the case, then we have our proof.
The same program, however, could be used just as well to refute the dual statement ``11 is a composite number''.
So in this case, the same construction would serve for proving one statement as well as for refuting the dual statement.
Or to take a `real-life' example, let us suppose we are in court and a video is shown recording person X shooting person Z, while it is person Y being in the dock.
Given that the video tape has been checked by experts for authenticity, it is clear in lighting, etc., this video would probably be taken as a refutation of the  court's charges that Y is the murderer of Z, or, to put it in a bit odd way in natural language, it would be taken as proof that Y is a non-murderer of Z.\footnote{Of course, in our natural language we do not have a strict definition of a ``dual proposition'' but one can come up with intuitive examples as I tried to do here. ``(Dual) proposition'' and ``(dual) statement'' are used synonymously in this context.}

\noindent Similar suggestions, drawing on Nelson and L\'{o}pez-Escobar, have been made in the more recent literature as well, e.g. by \cite{Wansing2016b} and \cite{Ferguson2020} for Nelson's constructive logic with strong negation, \texttt{N4}, namely that a construction $c$ can be taken as a proof of $A$ iff $c$ is a disproof of $\sim A$ (and vice versa).
Ferguson \citeyearpar[p. 1507]{Ferguson2020} argues here that there can be a coextensionality between a verifier of one formula and a falsifier of another, which, however, would not entail their identity because their sense differs.
While I would agree with the latter, I think that we must distinguish here very carefully what we mean by `identity'.
Whereas Ferguson seems to understand it as `being the same on all levels', I would understand it in the Fregean way, for which Frege uses `='.
Identity between `$a$' and `$b$' means that they have the same denotation but not necessarily that they have the same sense.
The intuition that verifiers and falsifiers may, in certain settings, be coextensional while differing in sense, can be well captured within the here proposed system, though.
However, from a bilateralist point of view it is in my opinion preferable not to have strong negation as a primitive connective in the language since, in a way, it stands against the bilateralist idea that refutation (or denial, or rejection, etc.) is a concept \emph{prior} to negation.
To briefly explain my concerns here: In \texttt{2Int} you can have a negation, even two, namely the intuitionistic negation, defined by $A \rightarrow \bot$ and the dual intuitionistic negation, which we call \emph{co-negation}, defined by $\top \Yleft A$.
Since they are defined via implication and co-implication, which are manifestations of the two derivability relations in the object language, this seems to me in accordance with refutation (and here also proof) being the more primitive concept upon which negation is defined.
However, incorporating strong negation would mean to have a \emph{primitive} connective that is basically expressing exactly what is expressed by our derivability relations.
Firstly, I simply do not see the need for that.\footnote{Since we do have negations, an objection coming from a ``Frege-Geach-point'' angle \cite[see, e.g.,][]{Horwich2005}, that we need a negation in our language to express it in subclauses of sentences (where an interpretation as refutation would not suffice), does not seem to be a concern here. This question came up in a discussion about an earlier draft of this paper with Dave Ripley, whom I want to thank for helping to clarify my thoughts on this.} 
What is more, this would give strong negation a special place from the bilateralist PTS point of view as opposed to the other connectives, which do not express this relation between proofs and refutations. 
I do not deem this desirable if we have the bilateralist meaning-giving component already incorporated in the proof system via the derivability relations.
Finally, strong negation would be non-congruential in our system, which leads to problems from a PTS point of view when it comes to the question of uniqueness.\footnote{See \cite[p. 579f.]{Humberstone2011} and \citeyearpar[p. 183, 187]{Humberstone2020a} on this, or on the issue of uniqueness specifically in bilateralist systems, \citep{Ayhan2021a}. An example to show that strong negation would be non-congruential are the formulas $\sim(A \rightarrow B)$ and $A ~\wedge \sim B$, which are interderivable only w.r.t. $\Rightarrow^{+}$ and $\sim\sim(A \rightarrow B)$ and $\sim(A ~\wedge \sim B)$, which are interderivable only w.r.t. $\Rightarrow^{-}$.}

\noindent So, I think we should read the `$=$' used in Definition \ref{duality} for the mapping of terms to their dual terms in the Fregean way telling us that we should identify those terms in the sense that they refer to the same object.
Thus, if we take the denotation of derivations to be referred to by their end-term, in a bilateralist setting derivational constructions\footnote{I use this somewhat clumsy expression instead of ``proof objects'' in order to avoid sounding like falling back to the unilateralist picture of giving preference to proofs over dual proofs.} are not only identified modulo belonging to the same equivalence class induced by the set of conversions\footnote{I will pass over the question here whether $\eta$-conversions should be considered identity-preserving, too. The literature is divided with respect to that question. Martin-L\"of \citeyearpar[p. 100]{Martin1975}, for example, does not agree that $\eta$-conversions are identity-preserving. Prawitz \citeyearpar[p. 257]{Prawitz1971}, on the other hand, seems to lean towards it when he claims that it would seem ``unlikely that any interesting property of proofs is sensitive to differences created by an expansion". He does not make a clear decision on that, though. Wideb\"ack \citeyearpar{Wideback}, relating to results in the literature on the typed $\lambda$-calculus like \citep{Friedman1975} and \citep{Statman}, argues for $\beta$-$\eta$-equality to give the right account of identity of proofs and Girard \citeyearpar[p. 16]{Girard1989} does the same, although he also mentions that $\eta$-equations ``have never been given adequate status" compared to the $\beta$-equations. For our purposes here it suffices to go with the `safe' option of the conversions we defined here.} but also by them being connected via the duality function $d$.
Thus, in \texttt{N2Int}$_\lambda$ derivations that are denoted by terms belonging to the same equivalence class induced by $\beta$-, p-, and s-conversions and derivations denoted by their respective dual terms are identical, they have the same underlying derivational object.
The sense, however, must clearly be more fine-grained and thus, should not be identified over the duality function, just as it is not identified over equivalence induced by conversions\footnote{The exception are $\alpha$-conversions, i.e., renaming of bound variables, which does not induce a difference in sense.}: Remember, two derivations ending on different, though perhaps normal-form-equal, end-terms always differ in sense.
The reason for this is that the way the proof object is represented is taken to be essentially different.
In the context of comparing proofs and refutations it is the same: Although one can argue that the underlying derivational construction is the same, the way it is constructed is essentially different; in the one case by proving something and in the other by refuting something.

\noindent So, if we consider the following exemplary derivations:
\vspace{0.2cm}

\quad 
 \infer[\scriptstyle\rightarrow I]{(\lambda x.x)^{+}: \rho \rightarrow \rho}{[x^{+}: \rho]} 
 \quad \quad \quad
 \infer[\scriptstyle\rightarrow I]{(\lambda x.x)^{+}: \sigma \rightarrow \sigma}{[x^{+}:\sigma]}
 \quad \quad \quad 
 \infer[\scriptstyle\rightarrow I]{(\lambda y.y)^{+}: \sigma \rightarrow \sigma}{[y^{+}:\sigma]}

\vspace{0.2cm}
\quad 
 \infer=[\scriptstyle\Yleft I^{d}]{(\lambda x.x)^{-}: \rho \Yleft \rho}{\llbracket x^{-}:\rho \rrbracket} 
 \quad \quad \quad 
 \infer=[\scriptstyle\Yleft I^{d}]{(\lambda x.x)^{-}: \sigma \Yleft \sigma}{\llbracket x^{-}:\sigma \rrbracket}
\quad \quad \quad 
 \infer=[\scriptstyle\Yleft I^{d}]{(\lambda y.y)^{-}: \sigma \Yleft \sigma}{\llbracket y^{-}:\sigma \rrbracket}

\vspace{0.2cm}
\noindent In these cases the respective derivations on the vertical as well as on the diagonal axes (i.e., pairs of derivations positioned diagonally to one another) are different in sense but not in denotation since their end-terms can be obtained from each other by our duality function.
For this it does not matter that different formulas are derived because what we are interested in is not the denotation of the formulas but of the derivation, i.e., the structure of the construction is decisive here.
The same holds for derivations with not only the same denotation but also the \emph{same sense}, which we have for the three derivations on the respective horizontal axes.
Although the signs that are used differ from each other, this difference is negligible because when it comes to the meaning of \emph{derivations} (not formulas or propositions etc.), it should not make a difference which atomic formulas are chosen as long as the derived formula is structurally the same. 
In terms of type theory we can say that it makes no difference as long as the \emph{principal type} of the term, i.e., the most general type that can be assigned to a term, is the same.\footnote{For example, for the term $(\lambda x.x)^{+}$, its types could be $p \rightarrow p$, $q \rightarrow q$, $(p \rightarrow q) \rightarrow (p \rightarrow q)$, etc., while its principal type would be $A \rightarrow A$.}

So, as long as the principal types of all terms occurring within two derivations are the same, the sense cannot differ, although the signs \emph{actually occurring} in the derivations can be different.
This is also why I prefer to use a Curry-style typing over a Church-style typing. 
In the latter system each term is usually uniquely typed, i.e., we would get a collapse of signs and sense: Since the sense is constituted by the terms occurring in a derivation, a differently typed term would automatically lead to a different sense. 
What leads to a difference in sense in our system is a difference in the principal types of the terms or a difference in the polarities.
We can see here why we need the polarities for philosophical reasons: Without these all of the above derivations would not only be identical when it comes to their denotation but also when it comes to their sense, i.e., they would all be synonymous.
However, as argued above, I think it makes good sense to say that the \emph{way} of construction is essentially different when proving something vs. refuting something, i.e., the sense should be distinguished here. 
That the polarity makes a difference in sense can also be motivated by looking at substitution, which was one of Frege's motivations to make this distinction: In intensional contexts we cannot substitute expressions with a different sense \emph{salva veritate}.
Here, we cannot substitute same terms of different polarities for one another.
In the first derivation on the left above, e.g., we could not just substitute $x^{+}$ with $x^{-}$ and leave the rest unchanged.
The application of $\rightarrow_I$ would not be feasible.
Thus, derivations can be claimed to constitute an intensional context. 
This seems at least not inappropriate for the general idea of PTS, which is, as Schroeder-Heister \citeyearpar{sep-proof-theoretic-semantics}  puts it, ``intensional in spirit, as it is interested in proofs and not just provability''.\footnote{See also \citep{PSH2016} and \citep{Tranchini2021} about an intensional notion of harmony.}

\noindent I want to make two final remarks about possible concerns that might be raised.
First of all, it should be emphasized that identifying proofs and refutations does \emph{not} mean to ultimately retreat to unilateralism again just because we have only one underlying object.
To begin with, unilateralism means more than relying on one concept.
It rather means to favor a certain concept, more specifically a `positive' one, over the other `negative' one.
This does not happen here, though: both proofs and refutations are on a par, they
are just conceived of being different ways to do the same thing.
Furthermore, the identity between proofs and refutations is only stated for the denotation, not for the sense, though.
Thus, we still do have means to distinguish between proofs and refutations, i.e., there is not a complete collapse between these two concepts.
Secondly, it has been remarked and questioned whether we can truly speak of \emph{one} underlying account to distinguish sense and denotation of derivations given that there is apparently a fundamental difference between this account in a unilateralist vs. a bilateralist setting.
It is true that there is an asymmetry here because in the bilateralist setting we always have two senses of the same denotation which are on equal standing, unlike in the unilateralist case, where differences in sense often go along with non-normality vs. normality.
But first of all, that is not necessarily so: we could have two non-normal derivations (in a unilateralist setting) reducing to the same normal form and thus, we would have two different senses without one being the `prior' one.
Moreover, I do not see why this is a worry about this still being the same account: unilateralism and bilateralism \emph{are} fundamentally different so that the applications of the framework to these settings also delivers different outcomes does not seem surprising to me.

\section{Conclusion and Outlook}

In this paper I established an extension of the $\lambda$-calculus with which a natural deduction system for the logic \texttt{2Int}, containing proofs and refutations, can be annotated.
For this system, called $\lambda^{2Int}$, I proved certain properties, which are typically considered important for $\lambda$-calculi, such as subject reduction.
Furthermore, using a duality function for terms and types, I established and proved a Dualization Theorem, which states that for every proof (resp. for every refutation) of a formula, a refutation (resp. a proof) of the dual formula can be given in this system. 
Using the term annotations to make this explicit, I argued then that in this system proofs and refutations should be identified when they are connected by the duality function, since the underlying construction of the derivations is fundamentally the same.
In a Fregean manner of distinguishing sense and denotation, then, proofs and refutations can be seen to have the same denotation, while, being presented in a different way, having a different sense.  
Thus, we have a, from a bilateralist point of view, very desirable equality between proofs and refutations.
Neither is reduced to the other but rather both are considered to be on equal footing, since they are simply different ways to represent the same object.

\noindent In  \citep{Ayhan2024b} my aim is to apply $\lambda^{2Int}$ to the sequent calculus system \texttt{SC2Int} \citep{Ayhan2021a} in order to see the results of comparing corresponding derivations between natural deduction and sequent calculus concerning their sense and denotation.
Finally, as was noted by an anonymous referee, there may also exist potential applications of this paper's content in the field of theoretical computer science. 
The work by Zeilberger \citeyearpar{Zeilberger2008, Zeilberger2009}, for example, shows some similiarities: Not only are the ideas of Gentzen, Prawitz, Dummett and Martin-Löf on proof-theoretic semantics discussed as a motivation to work in a bilateralist setting with direct proofs and direct refutations (and, depending on those, justified proofs and justified refutations), but also, the Curry-Howard correspondence is used to interpret a so-called \emph{polarized} logic.\footnote{I would like to thank the referee for making me aware of these specific relations to computer science.}

\medskip
\bibliographystyle{apacite}
\bibliography{ReferencesGeneral}

\appendix
\section*{Appendix}
\subsection{Definition of compatibility}

\begin{definition} \label{Compatibility}
A binary relation $\mathcal{R}$ on \texttt{Term$_{2Int}$} is \emph{compatible} iff it satisfies the following clauses for all $t, r, s, u \in$ \texttt{Term$_{2Int}$}:
\begin{enumerate}
\item  If $t \mathcal{R} r$, then $abort(t^{*})^{\dagger} \mathcal{R} abort(r^{*})^{\dagger}$.
\item If $t \mathcal{R} r$, then $\langle t^{*},s^{*} \rangle^{*} \mathcal{R} \langle r^{*},s^{*} \rangle^{*}$.
\item If $t \mathcal{R} r$, then $\langle s^{*},t^{*} \rangle^{*} \mathcal{R} \langle s^{*},r^{*} \rangle^{*}$.
\item If $t \mathcal{R} r$, then $inl(t^{*})^{*}\mathcal{R}inl(r^{*})^{*}$.
\item If $t \mathcal{R} r$, then $inr(t^{*})^{*}\mathcal{R}inr(r^{*})^{*}$.
\item If $t \mathcal{R} r$, then $(\lambda x^{*}.t^{*})^{*} \mathcal{R} (\lambda x^{*}.r^{*})^{*}$, for all variables $x$.
\item If $t \mathcal{R} r$, then $\{ t^{+},s^{-} \}^{*}\mathcal{R}\{ r^{+},s^{-} \}^{*}$.
\item If $t \mathcal{R} r$, then $\{ s^{+},t^{-} \}^{*}\mathcal{R}\{ s^{+},r^{-} \}^{*}$.
\item If $t \mathcal{R} r$, then $fst(t^{*})^{*}\mathcal{R}fst(r^{*})^{*}$.
\item If $t \mathcal{R} r$, then $snd(t^{*})^{*}\mathcal{R}snd(r^{*})^{*}$.
\item If $t \mathcal{R} r$, then $\texttt{case}~ t^{*}\{x^{*}.s^{\dagger} | y^{*}.u^{\dagger}\}^{\dagger}\mathcal{R} \texttt{case}~ r^{*}\{x^{*}.s^{\dagger} | y^{*}.u^{\dagger}\}^{\dagger}$, for all variables $x, y$.
\item If $t \mathcal{R} r$, then $\texttt{case}~ s^{*}\{x^{*}.t^{\dagger} | y^{*}.u^{\dagger}\}^{\dagger}\mathcal{R} \texttt{case}~ s^{*}\{x^{*}.r^{\dagger} | y^{*}.u^{\dagger}\}^{\dagger}$, for all variables $x, y$.
\item If $t \mathcal{R} r$, then $\texttt{case}~ s^{*}\{x^{*}.u^{\dagger} | y^{*}.t^{\dagger}\}^{\dagger}\mathcal{R} \texttt{case}~ s^{*}\{x^{*}.u^{\dagger} | y^{*}.r^{\dagger}\}^{\dagger}$, for all variables $x, y$.
\item If $t \mathcal{R} r$, then $App(t^{*}, s^{*})^{*} \mathcal{R} App(r^{*}, s^{*})^{*}$.
\item If $t \mathcal{R} r$, then $App(s^{*}, t^{*})^{*} \mathcal{R} App(s^{*}, r^{*})^{*}$.
\item If $t \mathcal{R} r$, then $\pi_1(t^{*})^{\dagger}\mathcal{R}\pi_1(r^{*})^{\dagger}$.
\item If $t \mathcal{R} r$, then $\pi_2(t^{*})^{\dagger}\mathcal{R}\pi_2(r^{*})^{\dagger}$.
\end{enumerate}
 
\end{definition}

\subsection{Proof of Dualization Theorem}

\begin{proof}[Proof of Dualization Theorem cont.]
\vspace{0.1cm}

If $(\Gamma; \Delta) \Rightarrow^{+} inl(t^{+})^{+}: A \vee B$, resp. \\$(\Gamma; \Delta) \Rightarrow^{-} inl(t^{-})^{-}: A \wedge B$, is of height $n+1$, then (by Generation Lemma 6.1, resp. 5.4) we have $(\Gamma; \Delta)\Rightarrow^{+} t^{+}:A$, resp. $(\Gamma; \Delta)\Rightarrow^{-} t^{-}:A$ with height at most $n$. 

\noindent Then by inductive hypothesis $(d(\Delta); d(\Gamma))\Rightarrow^{-} d(t^{+}): d(A)$, resp.\\ $(d(\Delta); d(\Gamma))\Rightarrow^{+} d(t^{-}): d(A)$ are of height at most $n$ as well.  

\noindent By application of $\wedge I^{d}_{1}$, resp. $\vee I_{1}$, we can construct a derivation of height $n+1$ s.t. $(d(\Delta); d(\Gamma))\Rightarrow^{-} inl(d(t^{+}))^{-}: d(A) \wedge d(B)$, resp. $(d(\Delta); d(\Gamma))\Rightarrow^{+} inl(d(t^{-}))^{+}: d(A) \vee d(B)$.
By our definition of dual terms $d(inl(t^{*})^{*}) = inl(d(t^{*}))^{d}$.
The same holds for the two cases of $inr(t^{*})^{*}$.

\vspace{0.1cm}

\noindent If $(\Gamma; \Delta) \Rightarrow^{+} (\lambda x^{+}.t^{+})^{+}: A \rightarrow B$ is of height $n+1$, then (by Generation Lemma 3.1) we have $(\Gamma, x^{+}:A; \Delta)\Rightarrow^{+} t^{+}:B$ with height at most $n$. 

\noindent Then by inductive hypothesis $(d(\Delta); d(\Gamma), x^{-}:d(A))\Rightarrow^{-} d(t^{+}): d(B)$ is of height at most $n$ as well.  

\noindent By application of $\Yleft I^{d}$ we can construct a derivation of height $n+1$ s.t. \\$(d(\Delta); d(\Gamma))\Rightarrow^{-} (\lambda x^{-}.d(t^{+}))^{-}:d(B) \Yleft d(A)$. 
By our definition of dual terms $d((\lambda x^{+}.t^{+})^{+}) = (\lambda x^{-}.d(t^{+}))^{-}$.
\vspace{0.1cm}

\noindent If $(\Gamma; \Delta) \Rightarrow^{-} (\lambda x^{-}.t^{-})^{-}: A \Yleft B$ is of height $n+1$, then (by Generation Lemma 4.4) we have $(\Gamma; \Delta, x^{-}:B)\Rightarrow^{-} t^{-}:A$ with height at most $n$. 

\noindent Then by inductive hypothesis $(d(\Delta), x^{+}:d(B); d(\Gamma))\Rightarrow^{+} d(t^{-}): d(A)$ is of height at most $n$ as well.  

\noindent By application of $\rightarrow I$ we can construct a derivation of height $n+1$ s.t. $(d(\Delta); d(\Gamma))\Rightarrow^{+} (\lambda x^{+}.d(t^{-}))^{+}:d(B) \rightarrow d(A)$. 
By our definition of dual terms $d((\lambda x^{-}.t^{-})^{-}) = (\lambda x^{+}.d(t^{-}))^{+}$.

\vspace{0.1cm}

\noindent If $(\Gamma; \Delta) \Rightarrow^{-} \{ s^{+}, t^{-}\}^{-}: A \rightarrow B$, resp. $(\Gamma; \Delta) \Rightarrow^{+} \{ s^{+}, t^{-}\}^{+}: A \Yleft B$ is of height $n+1$, then (by Generation Lemma 3.3, resp. 4.1) for $\Gamma = \Gamma' \cup \Gamma''$ and $\Delta = \Delta' \cup \Delta''$  we have $(\Gamma'; \Delta')\Rightarrow^{+} s^{+}:A$ and $(\Gamma''; \Delta'')\Rightarrow^{-} t^{-}:B$ with height at most $n$. 

\noindent Then by inductive hypothesis $(d(\Delta'); d(\Gamma'))\Rightarrow^{-} d(s^{+}): d(A)$ and \\$(d(\Delta''); d(\Gamma''))\Rightarrow^{+} d(t^{-}): d(B)$ are of height at most $n$ as well.  

\noindent By application of $\Yleft I$, resp. $\rightarrow I^{d}$, we can construct a derivation of height $n+1$ s.t. $(d(\Delta); d(\Gamma))\Rightarrow^{+} \{ d(t^{-}), d(s^{+})\}^{+}: d(B) \Yleft d(A)$, resp. $(d(\Delta); d(\Gamma))\Rightarrow^{-} \{ d(t^{-}), d(s^{+})\}^{-}: d(B) \rightarrow d(A)$. 
By our definition of dual terms \\$d(\{ s^{+}, t^{-}\}^{*}) = \{ d(t^{-}), d(s^{+})\}^{d}$.
\vspace{0.1cm}

\noindent If $(\Gamma; \Delta) \Rightarrow^{+} fst(t^{+})^{+}:A$ is of height $n+1$, then (by Generation Lemma 5.2) we have $(\Gamma; \Delta) \Rightarrow^{+} t^{+}:A \wedge B$ with height at most $n$. 

\noindent Then by inductive hypothesis $(d(\Delta); d(\Gamma))\Rightarrow^{-} d(t^{+}): d(A) \vee d(B)$ are of height at most $n$ as well.  

\noindent By application of $\vee E^{d}_{1}$ we can construct a derivation of height $n+1$ s.t.\\ $(d(\Delta); d(\Gamma))\Rightarrow^{-} fst(d(t^{+}))^{-}: d(A)$ .
By our definition of dual terms \\$d(fst(t^{+})^{+}) = fst(d(t^{+}))^{-}$.
\vspace{0.1cm}

\noindent If $(\Gamma; \Delta) \Rightarrow^{+} snd(t^{+})^{+}: B$ is of height $n+1$, then (by Generation Lemma 5.3) we have $(\Gamma; \Delta)\Rightarrow^{+} t^{+}:A \wedge B$ with height at most $n$. 

\noindent Then by inductive hypothesis $(d(\Delta); d(\Gamma))\Rightarrow^{-} d(t^{+}): d(A)  \vee d(B)$ is of height at most $n$ as well.  

\noindent By application of $\vee E^{d}_{2}$ we can construct a derivation of height $n+1$ s.t. \\$(d(\Delta); d(\Gamma))\Rightarrow^{-} snd(d(t^{+}))^{-}: d(B)$.
By our definition of dual terms \\ $d(snd(t^{+})^{+}) = snd(d(t^{+}))^{-}$.

\vspace{0.1cm}

\noindent If $(\Gamma; \Delta) \Rightarrow^{-} fst(t^{-})^{-}: A$ is of height $n+1$, then (by Generation Lemma 6.5) we have $(\Gamma; \Delta)\Rightarrow^{-} t^{-}:A \vee B$ with height at most $n$. 

\noindent Then by inductive hypothesis $(d(\Delta); d(\Gamma))\Rightarrow^{+} d(t^{-}): d(A)  \wedge d(B)$ is of height at most $n$ as well.  

\noindent By application of $\wedge E_{1}$ we can construct a derivation of height $n+1$ s.t. \\$(d(\Delta); d(\Gamma))\Rightarrow^{+} fst(d(t^{-}))^{+}: d(A)$.
By our definition of dual terms \\ $d(fst(t^{-})^{-}) = fst(d(t^{-}))^{+}$.

\vspace{0.1cm}
\noindent If $(\Gamma; \Delta) \Rightarrow^{-} snd(t^{-})^{-}:B$ is of height $n+1$, then (by Generation Lemma 6.6) we have $(\Gamma; \Delta) \Rightarrow^{-} t^{-}:A \vee B$ with height at most $n$. 

\noindent Then by inductive hypothesis $(d(\Delta); d(\Gamma))\Rightarrow^{+} d(t^{-}): d(A) \wedge d(B)$ are of height at most $n$ as well.  

\noindent By application of $\wedge E_{2}$ we can construct a derivation of height $n+1$ s.t.\\ $(d(\Delta); d(\Gamma))\Rightarrow^{+} snd(d(t^{-}))^{+}: d(B)$.
By our definition of dual terms \\$d(snd(t^{-})^{-}) = snd(d(t^{-}))^{+}$. 

\vspace{0.1cm}

\noindent If $(\Gamma; \Delta) \Rightarrow^{*} \texttt{case}~ r^{+}\{x^{+}.s^{*} | y^{+}.t^{*}\}^{*}: C$, resp.\\ $(\Gamma; \Delta) \Rightarrow^{*} \texttt{case}~ r^{-}\{x^{-}.s^{*} | y^{-}.t^{*}\}^{*}: C$ is of height $n+1$, then (by Generation Lemma 6.3, resp. 5.6)  for $\Gamma = \Gamma' \cup \Gamma'' \cup \Gamma'''$ and $\Delta = \Delta' \cup \Delta'' \cup \Delta'''$ we have $(\Gamma'; \Delta')\Rightarrow^{+} r^{+}: A \vee B$, $(\Gamma'', x^{+}:A; \Delta'')\Rightarrow^{*} s^{*}:C$ and $(\Gamma''', y^{+}:B; \Delta''')\Rightarrow^{*} t^{*}:C$, resp. $(\Gamma'; \Delta')\Rightarrow^{-} r^{-}: A \wedge B$, $(\Gamma''; \Delta'', x^{-}:A)\Rightarrow^{*} s^{*}:C$ and $(\Gamma'''; \Delta''', y^{-}:B)\Rightarrow^{*} t^{*}:C$ with height at most $n$. 

\noindent Then by inductive hypothesis $(d(\Delta'); d(\Gamma'))\Rightarrow^{-} d(r^{+}): d(A) \wedge d(B)$,\\ $(d(\Delta''); d(\Gamma''), x^{-}:d(A))\Rightarrow^{d} d(s^{*}): d(C)$ and $(d(\Delta'''); d(\Gamma'''), y^{-}:d(B))\Rightarrow^{d} d(t^{*}): d(C)$, resp. $(d(\Delta'); d(\Gamma'))\Rightarrow^{+} d(r^{-}): d(A) \vee d(B)$, $(d(\Delta''), x^{+}:d(A); d(\Gamma''))$ $\Rightarrow^{d} d(s^{*}): d(C)$ and $(d(\Delta'''), y^{+}:d(B); d(\Gamma'''))\Rightarrow^{d} d(t^{*}): d(C)$ are of height at most $n$ as well.
  
\noindent By application of $\wedge E^{d}$, resp. $\vee E$, we can construct a derivation of height $n+1$ s.t. $(d(\Delta); d(\Gamma)) \Rightarrow^{d} \texttt{case}~ d(r^{+})\{x^{-}.d(s^{*}) | y^{-}.d(t^{*})\}^{d}: d(C)$, resp. $(d(\Delta); d(\Gamma))$\\$ \Rightarrow^{d} \texttt{case}~ d(r^{-})\{x^{+}.d(s^{*}) | y^{+}.d(t^{*})\}^{d}: d(C)$. 
By our definition of dual terms $d(\texttt{case}~ r^{+}\{x^{+}.s^{*} | y^{+}.t^{*}\}^{*}) = \texttt{case}~ d(r^{+})\{x^{-}.d(s^{*}) | y^{-}.d(t^{*})\}^{d}$ and \\$d(\texttt{case}~ r^{-}\{x^{-}.s^{*} | y^{-}.t^{*}\}^{*}) = \texttt{case}~ d(r^{-})\{x^{+}.d(s^{*}) | y^{+}.d(t^{*})\}^{d}$.

\vspace{0.1cm}

\noindent If $(\Gamma; \Delta) \Rightarrow^{+} App(s^{+}, t^{+})^{+}: B$ is of height $n+1$, then (by Generation Lemma 3.2)  for $\Gamma = \Gamma' \cup \Gamma''$ and $\Delta = \Delta' \cup \Delta''$ we have $(\Gamma'; \Delta')\Rightarrow^{+} s^{+}:A \rightarrow B$ and $(\Gamma''; \Delta'')\Rightarrow^{+} t^{+}:A$ with height at most $n$. 

\noindent Then by inductive hypothesis $(d(\Delta'); d(\Gamma'))\Rightarrow^{-} d(s^{+}): d(B) \Yleft d(A)$ and \\$(d(\Delta''); d(\Gamma''))\Rightarrow^{-} d(t^{+}): d(A)$ is of height at most $n$ as well.  

\noindent By application of $\Yleft E^{d}$ we can construct a derivation of height $n+1$ s.t.\\ $(d(\Delta); d(\Gamma))\Rightarrow^{-} App(d(s^{+}), d(t^{+}))^{-}:d(B)$. 
By our definition of dual terms $d(App(s^{+}, t^{+})^{+}) = App(d(s^{+}), d(t^{+}))^{-}$.
\vspace{0.1cm}

\noindent If $(\Gamma; \Delta) \Rightarrow^{-} App(s^{-}, t^{-})^{-}: B$ is of height $n+1$, then (by Generation Lemma 4.5) for $\Gamma = \Gamma' \cup \Gamma''$ and $\Delta = \Delta' \cup \Delta''$  we have $(\Gamma'; \Delta')\Rightarrow^{-} s^{-}:B \Yleft A$ and $(\Gamma''; \Delta'')\Rightarrow^{-} t^{-}:A$ with height at most $n$. 

\noindent Then by inductive hypothesis $(d(\Delta'); d(\Gamma'))\Rightarrow^{+} d(s^{-}): d(A) \rightarrow d(B)$ and $(d(\Delta''); d(\Gamma''))\Rightarrow^{+} d(t^{-}): d(A)$ is of height at most $n$ as well. 
 
\noindent By application of $\rightarrow E$ we can construct a derivation of height $n+1$ s.t. $(d(\Delta); d(\Gamma))\Rightarrow^{+} App(d(s^{-}), d(t^{-}))^{+}:d(B)$. 
By our definition of dual terms $d(App(s^{-}, t^{-})^{-}) = App(d(s^{-}), d(t^{-}))^{+}$.

\vspace{0.1cm}

\noindent If $(\Gamma; \Delta) \Rightarrow^{+} \pi_1(t^{-})^{+}: A$ is of height $n+1$, then (by Generation Lemma 3.4) we have $(\Gamma; \Delta) \Rightarrow^{-} t^{-}:A \rightarrow B$ with height at most $n$. 

\noindent Then by inductive hypothesis $(d(\Delta); d(\Gamma))\Rightarrow^{+} d(t^{-}): d(B) \Yleft d(A)$ is of height at most $n$ as well.  

\noindent By application of $\Yleft E_{2}$ we can construct a derivation of height $n+1$ \\s.t. $(d(\Delta); d(\Gamma))\Rightarrow^{-} \pi_2(d(t^{-}))^{-}: d(A)$.
By our definition of dual terms  $d(\pi_1(t^{-})^{+})$\\ $= \pi_2(d(t^{-}))^{-}$.

\vspace{0.1cm}

\noindent If $(\Gamma; \Delta) \Rightarrow^{+} \pi_1(t^{+})^{+}: A$ is of height $n+1$, then (by Generation Lemma 4.2) we have $(\Gamma; \Delta) \Rightarrow^{+} t^{+}:A \Yleft B$ with height at most $n$.

\noindent Then by inductive hypothesis $(d(\Delta); d(\Gamma))\Rightarrow^{-} d(t^{+}): d(B) \rightarrow d(A)$ is of height at most $n$ as well. 

\noindent By application of $\rightarrow E^{d}_{2}$ we can construct a derivation of height $n+1$ s.t. $(d(\Delta); d(\Gamma))\Rightarrow^{-} \pi_2(d(t^{+}))^{-}: d(A)$.
By our definition of dual terms  $d(\pi_1(t^{+})^{+})$\\ $= \pi_2(d(t^{+}))^{-}$.

\vspace{0.1cm}

\noindent If $(\Gamma; \Delta) \Rightarrow^{-} \pi_2(t^{+})^{-}: A$ is of height $n+1$, then (by Generation Lemma 4.3) we have $(\Gamma; \Delta)\Rightarrow^{+} t^{+}:B \Yleft A$ with height at most $n$. 

\noindent Then by inductive hypothesis $(d(\Delta); d(\Gamma))\Rightarrow^{-} d(t^{+}): d(A) \rightarrow d(B)$ is of height at most $n$ as well.  

\noindent By application of $\rightarrow E^{d}_{1}$ we can construct a derivation of height $n+1$ s.t. $(d(\Delta); d(\Gamma))\Rightarrow^{+} \pi_1(d(t^{+}))^{+}: d(A)$.
By our definition of dual terms  $d(\pi_2(t^{+})^{-}) = \pi_1(d(t^{+}))^{+}$.

\vspace{0.1cm}

\noindent If $(\Gamma; \Delta) \Rightarrow^{-} \pi_2(t^{-})^{-}: A$ is of height $n+1$, then (by Generation Lemma 3.5) we have $(\Gamma; \Delta)\Rightarrow^{-} t^{-}:B \rightarrow A$ with height at most $n$. 

\noindent Then by inductive hypothesis $(d(\Delta); d(\Gamma))\Rightarrow^{+} d(t^{-}): d(A) \Yleft d(B)$ is of height at most $n$ as well.

\noindent By application of $\Yleft E_{1}$ we can construct a derivation of height $n+1$ s.t.\\ $(d(\Delta); d(\Gamma))\Rightarrow^{+} \pi_1(d(t^{-}))^{+}: d(A)$.
By our definition of dual terms  $d(\pi_2(t^{-})^{-}) = \pi_1(d(t^{-}))^{+}$.
\end{proof}

\end{document}